\documentclass[aps,pra,10pt,twocolumn,superscriptaddress,showpacs,preprintnumbers,amsmath,amssymb, longbibliography]{revtex4-1}

\usepackage{graphicx}
\usepackage{color}
\usepackage{dcolumn}
\usepackage{times}
\usepackage{bm}
\usepackage{verbatim}
\usepackage{mathbbol}
\usepackage{amsmath}
\usepackage{amssymb}
\usepackage{yfonts}
\usepackage{physics}
\usepackage{mathrsfs}
\usepackage{upgreek}
\usepackage[normalem]{ulem} 
\usepackage[caption=false]{subfig}
\usepackage{float}
\definecolor{dark_red}{rgb}{0.75,0,0}
\definecolor{dark_purple}{rgb}{0.75,0,0.75}
\definecolor{dark_blue}{rgb}{0,0,0.75}
\definecolor{dark_green}{rgb}{0,0.60,0}
\usepackage[colorlinks,citecolor=blue,linkcolor=dark_green]{hyperref}

\def\Xint#1{\mathchoice
   {\XXint\displaystyle\textstyle{#1}}%
   {\XXint\textstyle\scriptstyle{#1}}%
   {\XXint\scriptstyle\scriptscriptstyle{#1}}%
   {\XXint\scriptscriptstyle\scriptscriptstyle{#1}}%
   \!\int}
\def\XXint#1#2#3{{\setbox0=\hbox{$#1{#2#3}{\int}$}
     \vcenter{\hbox{$#2#3$}}\kern-.5\wd0}}

\def\dashint{\Xint-}

\newcommand{\NewStock}{\textsc{NewStock}}

\begin{document}

\title{Autoionizing Polaritons with the Jaynes-Cummings Model}

\author{Coleman Cariker}
\affiliation{Department of Physics, University of Central Florida, USA}
\author{S.~Yanez-Pagans}
\author{Nathan Harkema}%
\affiliation{Department of Physics, University of Arizona, Arizona, USA}
\author{Eva Lindroth}
\affiliation{Department of Physics, Stockholm University, Stockholm, Sweden, EU}
\author{Arvinder Sandhu}
\affiliation{Department of Physics, University of Arizona, Arizona, USA}
\author{Luca Argenti}
\affiliation{Department of Physics, University of Central Florida, USA}
\affiliation{CREOL, University of Central Florida, USA}
\email{luca.argenti@ucf.edu}

\date{\today}

\begin{abstract}
Intense laser pulses have the capability to couple resonances in the continuum, leading to the formation of a split pair of autoionizing polaritons. These polaritons can exhibit extended lifetimes due to interference between radiative and Auger decay channels. In this work, we show how an extension of the Jaynes-Cummings model to autoionizing states quantitatively reproduces the observed phenomenology. This extended model allows us to study how the dressing laser parameters can be tuned to control the ionization rate of the polariton multiplet. 
\end{abstract}
\maketitle

\section{\label{sec:intro} Introduction}
Attosecond pump-probe spectroscopy has gained prominence as a useful tool for probing and controlling ultrafast electronic dynamics in atoms~\cite{Berrah1996,Lukin2001,Fleischhauer2005,Pazourek2015,Goulielmakis2010,Mansson2014,Jimenez-Galan2014,Argenti2015,Kotur2016,Lin2013,Ossiander2016,Chu2010,Douguet2016,Barreau2016}, molecules~\cite{Xie2012,Joffre2007,Peng2015,Wang2021,Asenjo-Garcia2017,Haessler2010,Lepine2014,Huppert2016}, and condensed matter \cite{Sidiropoulos:22,Kruger2011,Fan:22}. 
The electronic continuum of polyelectronic atoms and molecules features localized transiently-bound states, which eventually emit an electron by Auger decay, a process driven by electron-electron correlation. These states, which are essential for the control of photoelectron emission processes~\cite{Lindroth2012}, appear as asymmetric peaks in the photo-absorption profile of atoms and molecules, due to interference between the direct-ionization path, and the radiative excitation of the metastable state followed by its autoionization~\cite{Fano_1965}. Quantum coherence, therefore, is an essential feature of these states, which manifests itself also in the temporal evolution of the electronic wavefunction, when interrogated with time resolved spectroscopies~\cite{GrusonScience2016}. External fields can modify the evolution of these metastable states, altering their spectral lineshapes and lifetimes~\cite{OttScience2013}.
The absorption line of bound or metastable states, when influenced by a dressing laser, varies with laser intensity and the proximity of other resonances. This variation includes phenomena such as AC Stark shift~\cite{Chini2012}, subcycle oscillations of the absorption intensity~\cite{Chini2013}, modification of the lineshape asymmetry~\cite{Yang2015}, and Autler-Townes (AT) splitting \cite{Ott_Nature2014,Zhao2017}. AT splitting, in particular, has been observed in the absorption profile of atoms both above~\cite{Wang2010,PfeifferPRA2012,Kobayashi2017,Miheli__2017} and below~\cite{Gaarde2011,WuGaardePRA2013,Zhao2017} the ionization threshold. In 1963, Jaynes and Cumming formulated a model for the AT splitting of two bound states in terms of quantized radiation states~\cite{Jaynes1963}. In this model, the two branches of an AT multiplet are regarded as entangled states of atomic and light configurations, known as polaritons~\cite{Galego2015,Zhang2019,Herrera2020}. 

Recent attosecond transient absorption measurements, corroborated by theoretical calculations, have revealed AT splitting of the $3s^{-1}4p$ resonance in argon, arising from its strong radiative coupling to the $3s^{-1}3d$ dark state~\cite{Harkema2022}. Given that these states are subject to autoionization decay, we refer to the AT branches as autoionizing polaritons (AIPs).
\begin{figure}[hbtp!]
\centering
\includegraphics[width=\columnwidth]{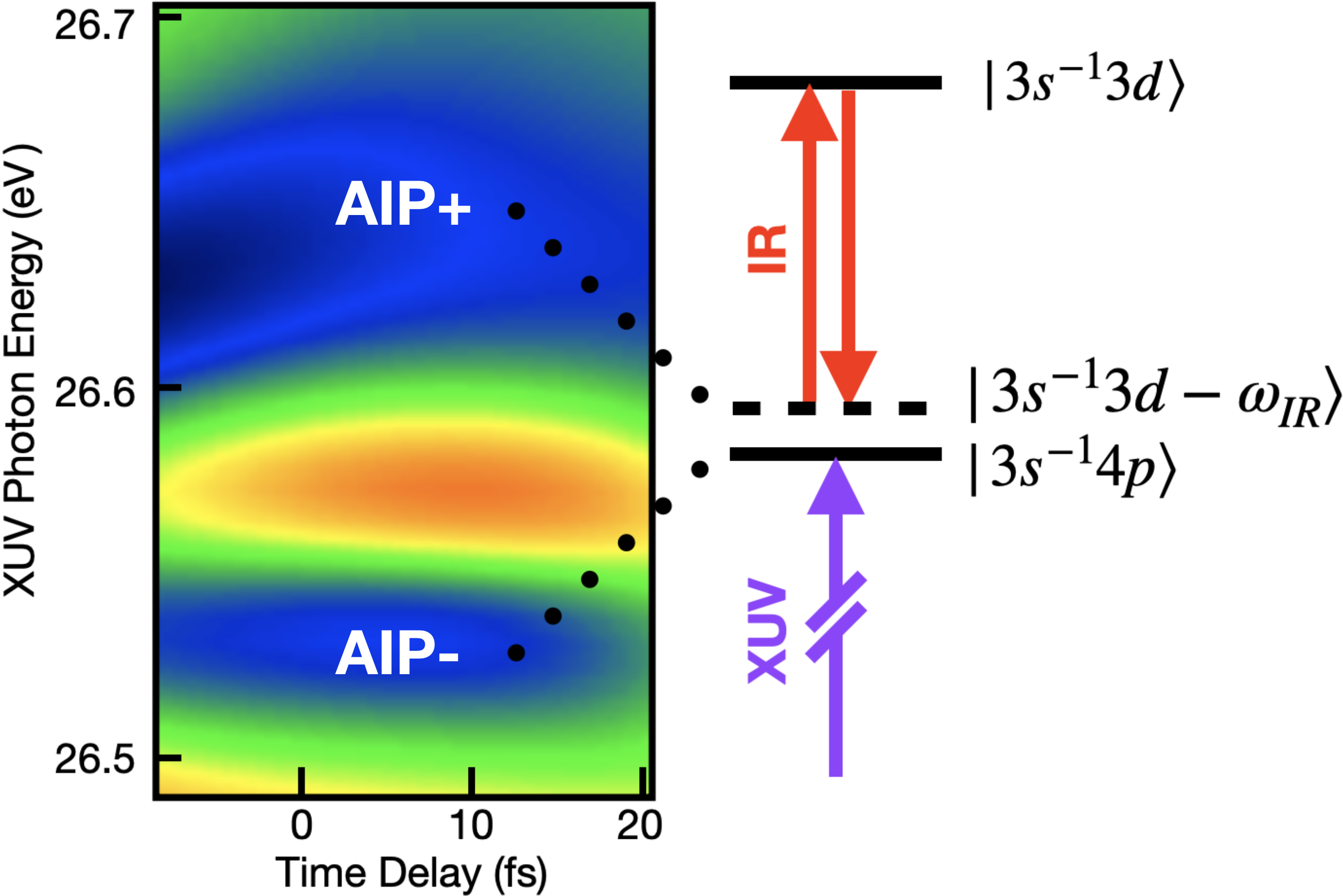}
\caption{\label{fig:ExampleATSplitting}\footnotesize Example of an \emph{ab initio} transient absorption spectrum in argon. Here, a moderately intense dressing pulse couples the $3s^{-1}4p$ and $3s^{-1}3d$ autoionizing states. The strong IR coupling splits the resonance into a pair of AIPs,
which are a linear combinations of the $3s^{-1}4p$ resonance and the $3s^{-1}3d-\omega_{IR}$ light-induced state.
}
\end{figure}
Figure~\ref{fig:ExampleATSplitting} exemplifies this splitting in the simulated extreme ultraviolet (XUV) attosecond transient absorption spectrum of the argon atom.
Autoionizing states (AIS), which are immersed in the continuum, are susceptible to photoionization even by low-energy photons like those from the infrared (IR) dressing field. External radiation fields, therefore, may be expected to reduce the lifetime of these states. Contrary to expectations, in 1982, Lambropolous and Z{\"o}ller posited that the Auger decay amplitude and the photoionization amplitude of a laser-dressed autoionizing state might interfere destructively, thereby stabilizing the state (see last paragraph in Sec.~VI of~\cite{Lambropoulos.PhysRevLett.49.1698}).

In the absence of any significant quantum interference between radiative- and auto-ionization amplitudes, the AIPs formed by resonant bright and dark states should have identical widths. However, the measurements presented in~\cite{Harkema2022} show that the widths of the two polaritons do differ, with one polariton being narrower than the original bright state, which is evidence of the coherent stabilization process predicted by Lambropoulos and Z\"oller. 
The interference between radiative and autoionization amplitudes enhancement in the AIP decay can be controlled via the laser parameters. AIPs, therefore, qualitatively differ from both bound-state polaritons, which do not have access to Auger decay channels, and overlapping autoionizing states, where radiation does not play any role. 

In this study, we reproduce some of the main features of overlapping autoionizing polaritons in the argon atom by means of \emph{ab initio} simulations. To explain the profiles observed in the simulations, we use Fano's approach to extend the Jaynes-Cummings (JC) to autoionizing states. Our model is capable of describing the optical excitation, from an initial ground state $|g\rangle$, of multiple autoionizing states coupled to each other by an external infrared driving field, and their subsequent decay to the continuum. It quantifies the interference between radiative and nonradiative decay pathways, illustrating how a laser can stabilize an autoionizing state, and it provides a functional form for the lineshape due to overlapping polaritons in transient absorption spectra, which was essential to extract the polariton parameters from the experiment~\cite{Harkema2022}. This contribution is a stepping stone for understanding coherent control of polaritonic states in matter, a key aspect of emerging technologies in quantum sensing~\cite{Peng2020}, photon trapping, and quantum information processing~\cite{Song2016}.

This paper is organized as follows. In Section~\ref{sec:abinitio}, we describe the \emph{ab initio} many-body calculations of argon's resonant attosecond transient-absorption using the \NewStock{} suite~\cite{Carette2013,Harkema2022}, demonstrating convergence of the results with a limited essential-state space. Section~\ref{sec:JCbasic} introduces an elementary version of the extended JC model, applicable to two non-overlapping resonances when the direct ionization amplitude from the ground state to the continuum is negligible. This simplified model is already capable of demonstrating the mechanism of destructive and constructive interference responsible for the stabilization and destabilization of the autoionizing polaritons observed in the \emph{ab initio} simulations and in the experiment. Section~\ref{sec:JCAdvanced} describes a more general extended JC model suited to describe multiple laser-coupled autoionizing states alongside a non-negligible direct ionization amplitude from the ground state to the continuum. This model is essential to derive the functional form for the profile of polaritonic multiplets, which is needed to extract the position and width of these states from experimental spectra. In section~\ref{sec:conclusion} we provide our conclusions.

\section{\emph{Ab initio} calculations}\label{sec:abinitio}
In this section we present the results of \emph{ab initio} simulations of the transient absorption spectrum of argon, focusing on the case where the $3s^{-1}4p$ resonance is excited by a weak XUV pump pulse, with a duration of few femtosecond, while being dressed by a moderately strong IR probe pulse. Depending on its central frequency, the IR pulse may bring in resonance the $3s^{-1}4p$ bright state with other dark or bright resonances by means of one-photon or two-photon transitions, thus causing the resonance peak to split into a pair of AIPs. 

For our study, we assume that both the pump and probe pulses are linearly polarized along the same quantization axis. To reproduce the observables measured in a realistic experiment, we compute the transient absorption spectrum both as a function of the pump-probe delay (keeping the IR intensity and frequency constant) and as a function of the IR frequency (at a fixed time delay). 

In our \emph{ab initio} atomic structure calculations, we employ the \NewStock{} suite of atomic codes, which expresses the wave function for the neutral system a time-dependent extended close-coupling (CC) expansion
\begin{equation}\label{eq:tdcc}
\begin{split}
\Psi(\boldsymbol{x};t) = \sum_{\Gamma}\Bigg[&\sum_{\alpha}\mathcal{A}\,\Phi_{\alpha}^\Gamma(\bar{\boldsymbol{x}};\hat{r}_{N_e},\zeta_{N_e})\,\varphi_{\alpha}^\Gamma(r_{N_e};t)\,+\\
&+\,\langle \boldsymbol{x} |\mathbf{K}^\Gamma\rangle\, \mathbf{c}^\Gamma(t)\Bigg],
\end{split}
\end{equation}
where the vectors $\boldsymbol{x}=(x_1,\,x_2,\,\ldots,\,x_{N_e})$ and $\bar{\boldsymbol{x}}=(x_1,\,x_2,\,\ldots,\,x_{N_e-1})$ contain the spatial and spin coordinates for the electrons in the neutral atom and in the ion, respectively, with $x_i=(\vec{r}_i,\zeta_i)$, and $\mathcal{A}=\frac{1}{N_e!}\sum_{\mathcal{P}\in\mathcal{S}_{N_e}}\mathrm{sgn}(\mathcal{P})\mathcal{P}$ is the antisymmetrizer. The index $\alpha$ identifies a partial-wave channel (PWC), i.e., a specific ionic state coupled to a photoelectron with definite orbital angular momentum to give rise to an atomic state with specified parity, $\Pi$, orbital angular momentum, $L$, spin, $S$, and their respective projections, $M$ and $\Sigma$, indicated collectively by the symbol  $\Gamma=(\Pi,L,S,M,\Sigma)$,
\begin{equation}
\begin{split}
\Phi_{\alpha}^\Gamma(\bar{\boldsymbol{x}};\hat{r}_{N_e},\zeta_{N_e}) &= \sum_{M_am}\sum_{\Sigma_a \sigma} C_{L_aM_a,\ell m}^{LM} C_{S_a\Sigma_a,\frac{1}{2}\sigma}^{S\Sigma}\times\\ 
&\times\phi_a(\bar{\boldsymbol{x}})\,Y_{\ell m}(\hat{r}_{N_e})\,{^2\chi_\sigma}(\zeta_{N_e}),
\end{split}
\end{equation}
where $C_{\ell_1 m_1, \ell_2 m_2}^{\ell_{12},m_{12}}$ are Clebsch-Gordan coefficients, $\phi_a(\bar{\boldsymbol{x}})$ is the state of the ion in channel $\alpha$, $Y_{\ell m}(\hat{r})$ are spherical harmonics, $^2\chi_\sigma(\zeta)=\langle \zeta|\sigma\rangle = \delta_{\sigma\zeta}$, with $\sigma,\,\zeta=\pm\frac{1}{2}$, are spin eigenfunctions, whereas $\varphi_\alpha^\Gamma(r;t)$ in \eqref{eq:tdcc}, is the photoelectron radial function, which can depend on time. Finally the row vector $|\mathbf{K}^\Gamma\rangle = (|K_1^\Gamma\rangle,\,|K_2^\Gamma\rangle,\,\ldots)$ specifies a set of configuration-state functions (CSFs), complementary to the PWC functions, called localized channel (LC), where $K_i^\Gamma$ is a given atomic CSF, e.g., $1s^22s^22p^63s^23p^54s({^1P^o})$, $1s^22s^22p^63s^23p^4({^3P^e})4s({^2P^e})4p({^1P^o})$, etc. 
The ionic functions $\phi_a(\bar{\boldsymbol{x}})= \langle \bar{\boldsymbol{x}} | \phi_a\rangle$ are themselves expressed in terms of a set of ionic CSFs, $|\bar{\mathbf{K}}\rangle$,
\begin{equation}
    |\phi_a\rangle = \sum_{\bar{\mathbf{K}}}| \bar{\mathbf{K}}^{\Gamma_a}\rangle\, C^{\Gamma_a}_{\bar{\mathbf{K}},a}.
\end{equation}
In this work, the ionic CSFs arise from the following shell occupations (for brevity, the inactive neon $1s^22s^22p^6$ core is not specified): $3s^23p^5$, $3s^23p^43d$, $3s^23p^44s$, $3s^23p^44p$, $3s^23p^33d^2$, $3s^23p^33d4s$, $3s^23p^33d4p$, $3s^23p^34s4p$, $3s^23p^34p^2$, $3s^23p^23d^3$, $3s^23p^23d^24s$, $3s3p^6$, $3s3p^53d$, $3s3p^54p$, $3s3p^43d^2$, $3s3p^43d4s$, $3s3p^43d4p$, and $3p^53d^2$, which give rise to 25 {$^2S^e$}, 40 {$^2P^o$}, 51 {$^2P^e$}, and 65 {$^2D^e$} ionic CSFs.  The orbitals and the ionic states are optimized using the Multi-configuration Hartree-Fock (MCHF), implemented in the \textsc{atsp2k} atomic-structure package~\cite{Froese-Fischer1997}, by minimizing the weighted average of the energy of the first $^2S^e$ and the first $^2P^o$ states of the ion, which have dominant configuration [Ar]$3s^{-1}$ and [Ar]$3p^{-1}$, respectively. 
To construct the PWCs used in \ref{eq:tdcc}, we couple the first two MCHF ions in each of the $^2P^o$, $^2S^e$, $^2D^e$, and $^2P^e$ ion symmetries, to photoelectrons with orbital angular momentum up to $\ell_{\mathrm{max}}=4$, resulting in  6, 12, 4, 14, 12, and 10 PWCs in the neutral-atom symmetries $^1S^e$, $^1P^o$, $^1P^e$, $^1D^e$, $^1F^o$, and $^1G^e$, respectively. The radial functions for the PWC photoelectrons are selected within the orthogonal complement to the ionic active orbitals of a space of radial B-splines~\cite{Carette2013,Argenti2006} with order 7, spanning the interval $r\in[0:R_{\textsc{BOX}}]$\,a.u., with $R_{\textsc{BOX}}=500$\,a.u. and an asymptotic spacing of 0.4~a.u. between consecutive nodes.  The LC in \eqref{eq:tdcc} consists of all the configurations formed by adding an electron in any of the active ionic orbitals $3s,\,3p,\,3d,\,4s,\,4p$ to any of the ionic CSFs. The size of the LCs in the various symmetries is: 193 for $^1S^e$, 435 for $^1P^o$, 314 for $^1P^e$, 509 for $^1D^e$, 449 for $^1F^o$, and 339 for $^1G^e$. 

The initial ground state of the system, $|g\rangle$, is obtained by diagonalizing the field-free electrostatic Hamiltonian $\hat{H}_0$ in the close-coupling basis. Formally, the subsequent evolution of the system under the influence of the external sequence of pump and probe pulses is determined by solving the time-dependent Schrödinger equation (TDSE), within the dipole approximation, in velocity gauge~\cite{Bransden2003},
\begin{equation}
i\hbar\partial_t |\Psi(t)\rangle = \left[\hat{H}_0+\frac{1}{c}\vec{A}(t)\cdot\hat{\vec{P}}\right]\,|\Psi(t)\rangle,\quad |\Psi(t_i)\rangle = |g\rangle,
\end{equation} 
where $\hat{\vec{P}}=\sum_{i=1}^{N_e}\hat{\vec{p}}_i$ is the total electron linear momentum and $\vec{A}(t) = \hat{z}\,A(t)$ is the vector potential of the external field. Gauss units~\cite{Jackson} and atomic units ($\hbar=1$, $m_e=1$, $q_e=-1$) are used throughout the paper unless specified otherwise. The absorption spectrum, or optical density (OD), in the XUV region is finally determined from the Fourier Transform of the dipole expectation value along the field polarization, $\tilde{P}(\omega) = \int e^{i\omega t}\, P(t)\,dt$, where $P(t)=\langle\Psi(t)|\hat{P}_z|\Psi(t)\rangle\,dt$, using the formula
\begin{equation}\label{eq:OD}
    \sigma(\omega)=-\frac{4\pi}{\omega}\Im\left[ \frac{\tilde{P}(\omega)}{\tilde{A}(\omega)}\right],
\end{equation}

In practice, the XUV components OD are largely insensitive to the portion of the wave function far from the nucleus. Furthermore, propagation in a finite box would eventually lead to unphysical reflections at the box boundaries which would manifest themselves as spurious noise in the spectrum, as the reflected wave packet returns to the interaction region. This circumstance indicates both that artificial reflections must be prevented and that the configuration space could be significantly reduced. We achieve both goals by introducing in the field-free Hamiltonian a complex absorption potential $V_{\textsc{CAP}}$, defined as
\begin{equation}\label{eq:CAPs}
\hat{V}_{\textsc{CAP}}=-i\, \Gamma_{\textsc{CAP}}\sum_{i=1}^{N_e}\theta(\hat{r}_i-R_{\textsc{CAP}})(\hat{r}_i-R_{\textsc{CAP}})^2,
\end{equation}
where $\Gamma_{\textsc{CAP}}$ and $R_{\textsc{CAP}}$ are real positive constants, and $\theta(x)=\int_{-\infty}^{x} dx' \delta(x')$ is the Heaviside step function. In this calculation, $R_{\textsc{CAP}}=R_{\textsc{BOX}}-50\,\mathrm{a.u.}=450\,\mathrm{a.u.}$, and $\Gamma_{\textsc{CAP}}=8\times 10^{-4}$. The CAP estinguishes the wavefunction before it reaches the boundary of the quantization box, without causing itself any appreciable reflections. 

The matrix representative in the CC basis $|\boldsymbol{\chi}\rangle=(|\chi_1\rangle,\,|\chi_2\rangle,\,\ldots)$ of the complex symmetric field-free Hamiltonian with the CAP, $\mathbf{H}_0+\mathbf{V}_{\textsc{CAP}} = \langle \boldsymbol{\chi}|\hat{H}_0+\hat{V}_{\textsc{CAP}}|\boldsymbol{\chi}\rangle$, is diagonalized, 
\begin{equation}
\mathbf{H}_0+\mathbf{V}_{\textsc{CAP}}=\mathbf{U}_{R}\,\tilde{\mathbf{E}}\,\mathbf{U}_L^\dagger,\qquad\tilde{\mathbf{E}}_{ij}=\tilde{E}_{i}\,\delta_{ij},
\end{equation}
where $\tilde{E}_i=\bar{E}_i-i\Gamma_i/2$ are complex eigenvalues with negative imaginary parts, $\bar{E}_i,\,\Gamma_i\in\mathbb{R}$, $\Gamma_i\geq0$. The matrices $\mathbf{U}_{R/L}$ are collections of right and left eigenvectors of the Hamiltonian, normalized such that $\mathbf{U}_L^\dagger\mathbf{U}_{R}=\mathbf{1}$. The CAP effectively enforces outgoing boundary conditions on the autoionizing states of the system, which in this calculation appear as isolated eigenvectors of the complex Hamiltonian. Therefore, for $r_i<R_{\textsc{CAP}}$, the autoionizing states obtained with this procedure are Siegert states of the atom~\cite{Siegert1939,Tolstikhin2006}. In the following, therefore, we will refer to the basis of the $\mathbf{U}_{R}$ eigenstates of the complex Hamiltonian as Siegert-state (SS) basis.
\begin{figure}
    \centering
    \includegraphics[width=\columnwidth]{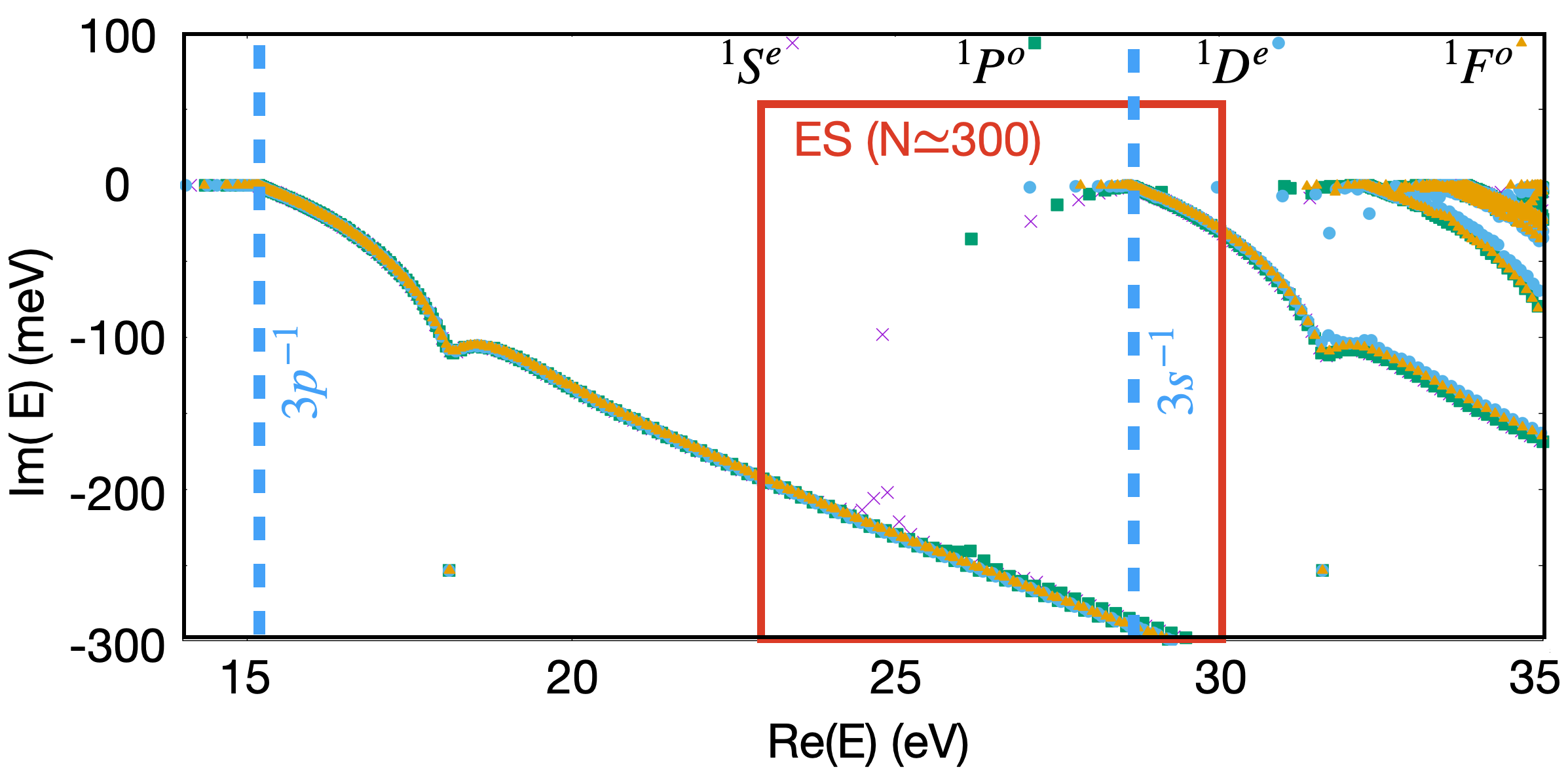}
    \caption{\label{ESBasis}Depiction of states retained from close-coupling calculations for the essential states basis, with symmetry $^1S^e$ (crosses, magenta online), $^1P^o$ (full square, green online), $^1D^e$ (full circles, cyan online), and  $^1F^o$ (full triangles, orange online).}
\end{figure}
Figure~\ref{ESBasis} shows a section of the complex spectrum of the quenched Hamiltonian in $^1S^e$, $^1P^o$, $^1D^e$, and $^1F^o$ symmetry. In the figure, we can clearly distinguish three main types of eigenvalues: bound states, with zero imaginary part, located below the first ionization threshold; discretized continuum states, branching off from the real axis at each channel's threshold $E_{\mathrm{th}}$ with increasingly negative imaginary parts; and autoionizing states, which correspond to isolated poles near the real axis, largely insensitive to the specific value of $\Gamma_{\textsc{CAP}}$ (as soon as it is large enough), and which here form Rydberg series converging to the ion shake-up thresholds. When the TDSE is propagated in the SS basis, we observed that the basis can be reduced to as little as 1\% of its full size, retaining only the states in the spectral region of interest --- e.g., those delimited by the red frame in Fig.~\ref{ESBasis} --- without affecting the quality of the simulated absorption spectrum. This remarkable finding reflects the circumstance that the transient absorption spectrum primarily probes the inner region of the wave function. We refer to this smaller set of state vectors as essential-state (ES) basis.

\begin{figure}
    \centering
    \includegraphics[width=\columnwidth]{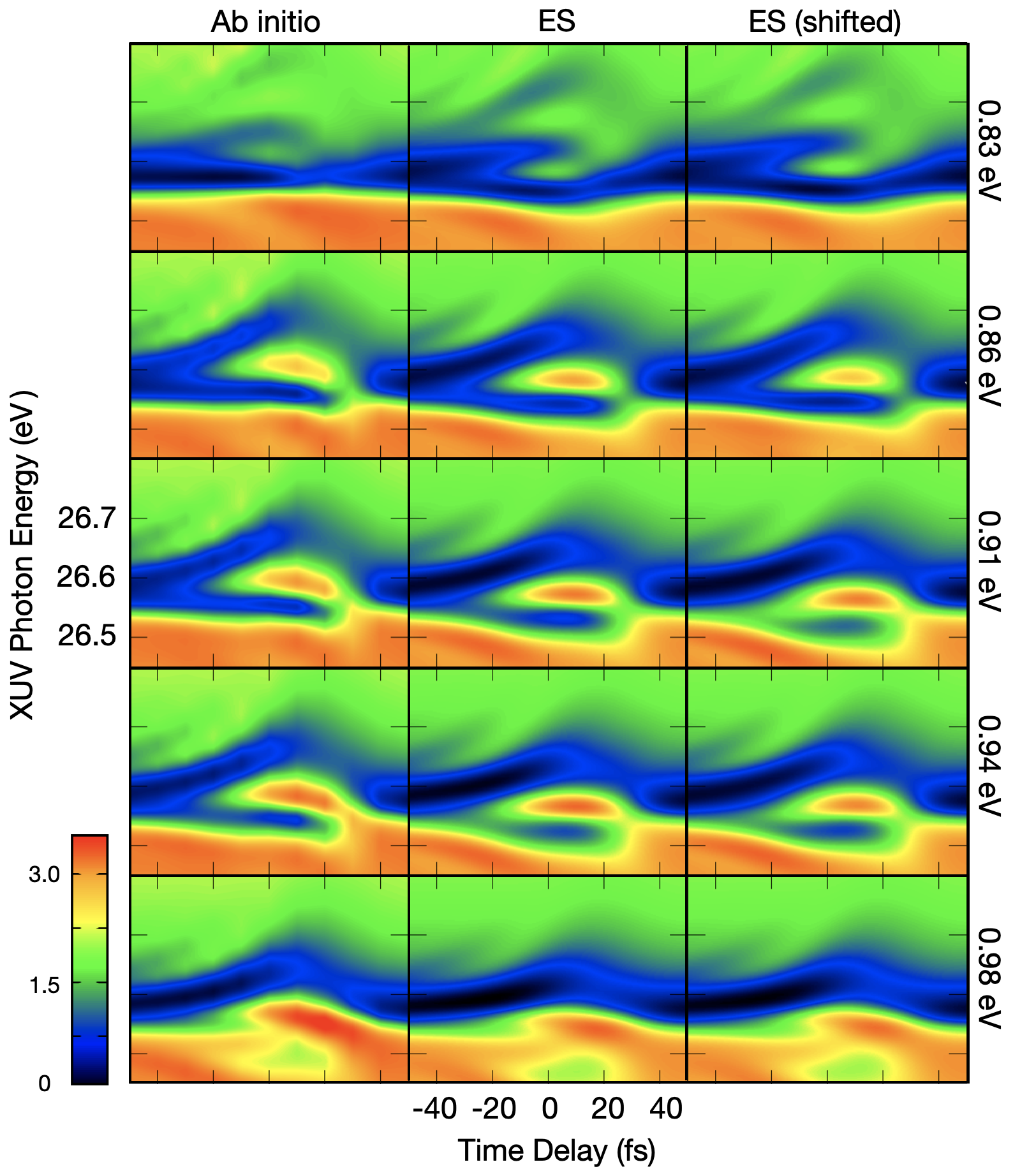}
    \caption{\label{AbES}Comparison of \emph{ab initio} calculation of the absorption spectrum as a function of time delay (left) at various IR energies with results from the essential states model. In the middle, resonance positions are kept as computed; on the right, they are shifted to match experimentally measured energies, with the IR energy adjusted to maintain constant $3s^{-1}4p-3s^{-1}3d$ detuning. Listed IR energies correspond to the unshifted calculations. In the top (bottom) row, the IR is tuned below (above) the resonant transition, positioning the light-induced states above (below) the $3s^{-1}4p$ resonance.}
\end{figure}

To illustrate the usefulness of the ES basis, we compare the transient-absorption spectra computed by solving the TDSE in both the full CC basis and in the ES basis. In this work, we employ a second-order split-exponential propagator for the wave function, represented in the SS (or ES) basis as $|\Psi(t)\rangle = \mathbf{U}_R\, \mathbf{c}(t)$,
\begin{equation}
\begin{split}
    \mathbf{c}(t+dt) =e^{-i\frac{dt}{2}\tilde{\mathbf{E}}}\,e^{-i dt/c A(t+\frac{dt}{2})\mathbf{P}_z}\,e^{-i\frac{dt}{2}\tilde{\mathbf{E}}}\mathbf{c}(t),
\end{split}
\end{equation}
where $\mathbf{P}_z = \mathbf{U}_L^\dagger \langle \boldsymbol{\chi} | \hat{P}_z |\boldsymbol{\chi}\rangle \mathbf{U}_R $ is the matrix representative, in the SS (ES) basis, of the velocity-gauge dipole operator along the field polarization.
While the complex field-free Hamiltonian can be diagonalized in the full CC basis, owing to its being block diagonal, $\langle\boldsymbol{\chi}^\Gamma | \hat{H}_0 + \hat{V}_{\textsc{CAP}} | \boldsymbol{\chi}^{\Gamma'\neq\Gamma}\rangle =0$, the dipole operator is normally too large to be diagonalized in the same basis, as it couples all the symmetries relevant to the simulation with each other.
In the full SS basis, therefore, the exponential $e^{-i dt/c \,A(t+\frac{dt}{2})\mathbf{P}_z}$ is estimated using an iterative Krylov solver~\cite{Youcef1989}, which requires several matrix-vector multiplications in the full CC basis at each time step. Since the ES basis can be two orders of magnitude smaller the full CC basis, on the other hand, the dipole operator can be diagonalized once and for all at the outset,
\begin{equation}
\mathbf{P}_z = \mathbf{W}_R \mathbf{\Lambda} \mathbf{W}^\dagger_L,\quad \mathbf{\Lambda}_{ij} = \Lambda_i \delta_{ij},\quad  \mathbf{W}^\dagger_L\mathbf{W}_R=\boldsymbol{1},
\end{equation}   
so that the propagation can be written as
\begin{equation}
\begin{split}
    \mathbf{c}(t+dt) =e^{-i\frac{dt}{2}\tilde{\mathbf{E}}}\,\mathbf{W}_R \,e^{-i \frac{dt}{c} A(t+\frac{dt}{2})\mathbf{\Lambda}}\, \mathbf{W}^\dagger_L e^{-i\frac{dt}{2}\tilde{\mathbf{E}}}\mathbf{c}(t),\nonumber
\end{split}
\end{equation}
which requires only two matrix-vector multiplications in a much smaller basis, accelerating the process by as much as four orders of magnitude. As shown in the first two columns of Fig.~\ref{AbES}, which is discussed in more detailed below, these two calculations yield nearly identical results.

Using the SS (ES) basis offers another marked advantage when simulating a realistic experiment. Since the energies of the autoionizing states are isolated complex eigenvalues, we can manually redefine their energies and widths to align them with the values known from the experiment, if available. Indeed, the \emph{ab initio} prediction of the resonance parameters may not be fully converged with respect to correlation, which can become an issue in case of observables that are extremely sensitive to the detuning between the external field frequency and the energy gap between radiatively coupled autoionizing states, which is the case here. Furthermore, we can selectively exclude individual autoionizing states from the simulation to verify their role in the pump-probe excitation process. This capability, therefore, is crucial for both identifying the mechanisms underlying specific spectral features and achieving quantitative agreement with experimental data. 

The need for adjusting the energies of the autoionizing states is exemplified by the calculated energies of the $3s^{-1}3d$, $3s^{-1}5s$, and $3s^{-1}6p$ resonances, which slightly differ from their experimental counterparts in terms of the energy distance from the $3s^{-1}3p$ state, as shown in table~\ref{ResPos}.  
\begin{table}[h!]
    \caption{Comparison between the theoretical (th) and experimental (exp) energy separation $\Delta E$ of certain relevant resonances from the $3s^{-1}4p$ state.}
    \label{ResPos}
\begin{ruledtabular}
    \begin{tabular}{ccc}
         AIS &  $\Delta E^{\mathrm{th}}$ (eV) & $\Delta E^{\mathrm{exp}}$ (eV) \\
         \hline \vspace{-3mm}\\
         $3s^{-1}3d$ & 0.91 & 0.89 \\
         $3s^{-1}5s$ & 0.92 & 0.93 \\
         $3s^{-1}6p$ & 1.82 & 1.90 \\
    \end{tabular}
\end{ruledtabular}
\end{table}
In our calculations, we adjust the positions of those resonances so that their detuning from the $3s^{-1}4p$ state matches the experimental conditions~\cite{Ke-Zun2003,Carette2013}, $\tilde{E}_i \mapsto \tilde{E}_i + \Delta {E}_{i}^{\mathrm{exp}} -  \Delta {E}_{i}^{\mathrm{exp}}$.  Ideally, we would also adjust the position of the $3s^{-1}4f$, but, to the best of our knowledge, there is no reported experimental value for this resonance in the literature.

Figure~\ref{AbES} presents a comparative analysis of the theoretical transient absorption spectrum of argon in the vicinity of the $3s^{-1}4p$ resonance, as a function of the pump-probe delay, for several values of the IR frequency, from positive (top row) to increasingly negative values of the detuning from the $3s^{-1}3d$ resonance, $\delta = E_{3s^{-1}3d}-E_{3s^{-1}4p}-\omega_{\textsc{IR}}$. These simulations were conducted using a 4~fs XUV pulse centered on the $3s^{-1}4p$ and a 50~fs IR pulse with a peak intensity of 40 GW/cm$^2$. The simulations were conducted in three different bases: i) the SS basis (first column), which includes the full close-coupling space, amounting to the order of $10^4$ states per symmetry; ii) the ES basis (second column), which encompasses a selective set of about $10^2$ essential states per symmetry; the ES-shifted basis (third column), which comprises the same states as the ES basis but adjusts the energy of the resonances listed in table~\ref{ResPos}. In the simulation in the ES-shifted basis, furthermore, the central frequency of the IR pulse is also modified so that the detuning $\delta$ is consistent across all calculations. 
In all the cases shown in Fig.~\ref{AbES}, a global shift of 0.405~eV is applied to the XUV photon energy to align the position of the $3s^{-1}4p$ with its experimentally observed value~\cite{Harkema2022}. To account for the finite spectrometer resolution in experiments, the simulation results are convoluted with a Gaussian function with 25\,meV full width at half maximum (FWHM).

As the IR is tuned to the $4p-3d$ transition (middle row), the $4p$ state splits into a pair of autoionizing polaritons due to the IR intensity and the strong coupling between the resonances. The polaritonic branches significantly differ in their widths, which are modulated by the time delay of the IR. As mentioned above, the comparison between the SS and the ES calculations in Figure~\ref{AbES} shows how excluding up to 99\% of the basis states does not significantly alter the simulation results. This demonstrates the robustness and efficiency of the ES approach in capturing the essential dynamics of the system. Stated otherwise, the time-dependent Hamiltonian for the present pump-probe pulses, when starting from the ground state, is rank sparse, and such sparsity is conveniently captured by the factorization of the Hamiltonian in the SS basis. Furthermore, the comparison between the ES and the shifted ES calculations highlights the paramount importance of the detuning between the IR dressing field and the energy separation of different resonances. As observed, when the detuning is consistent across calculations, the spectral shapes in the central and right columns are remarkably similar. This result shows that with the ES approach it is possible to match the frequency of the IR to its nominal experimental value while also preserving the same detuning as in the experiment.

\begin{figure}
    \centering
    \includegraphics[width=\columnwidth]{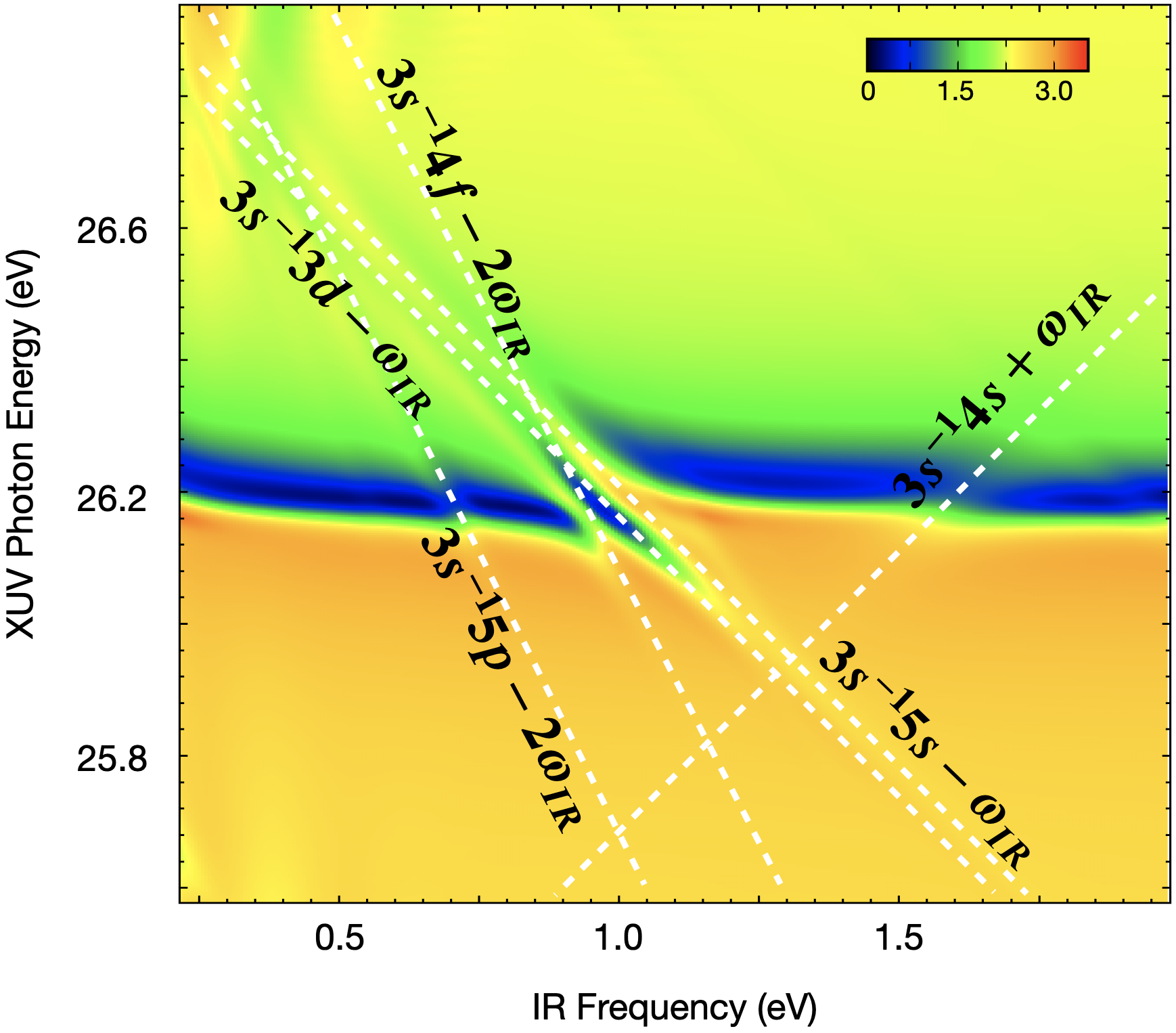}
    \caption{\emph{Ab initio} calculation of the absorption spectrum $\sigma(\omega_{\textsc{xuv}};\omega_{\textsc{ir}},I_{\textsc{ir}},\tau)$ at time delay $\tau=0$, as a function of the dressing laser frequency $\omega_{\textsc{IR}}$. The nominal position of the various LIS as a function of $\omega_{\textsc{IR}}$, $E_{3s^{-1}n\ell + n\gamma_{\textsc{IR}}}(\omega_{\textsc{IR}})=E_{3s^{-1}n\ell}+n\hbar\omega_{\textsc{IR}}$, are marked with white dashed lines and labelled according to their corresponding resonances. The figure clearly shows the avoided crossing between these LIS states and the bright $3s^{-1}4p$ window resonance.}
    \label{fig:FreqSweep}
\end{figure}
Now that we have ascertained that the ES-shifted basis can serve as a convenient model for the experiment, Fig.~\ref{fig:FreqSweep} shows the absorption spectrum $\sigma(\omega_{\textsc{xuv}};\omega_{\textsc{ir}},I_{\textsc{ir}},\tau)$ computed with this basis, as a function of the dressing-laser frequency. In this range of IR energies we can observe the effect of several one and two-photon couplings between the $3s^{-1}4p$ and other resonances in the continuum, which manifest themselves as avoided crossings. The nominal energies $E_{a}\pm n\omega_{\textsc{IR}}$ of the LIS corresponding to each dark state $a$ coupled to the $^1P^o$ symmetry by an $n$-photon transition, is indicated on the frequency map with a white dashed line. These lines intersect the main $3s^{-1}4p$ signal right where the avoided crossings appear. In particular, in the frequency interval around $\omega_{\textsc{IR}}\simeq$1\,eV, as many as three LIS overlap with the bright state, giving rise to a complex polaritonic multiplet.

\begin{figure}
    \centering
    \includegraphics[width=\columnwidth]{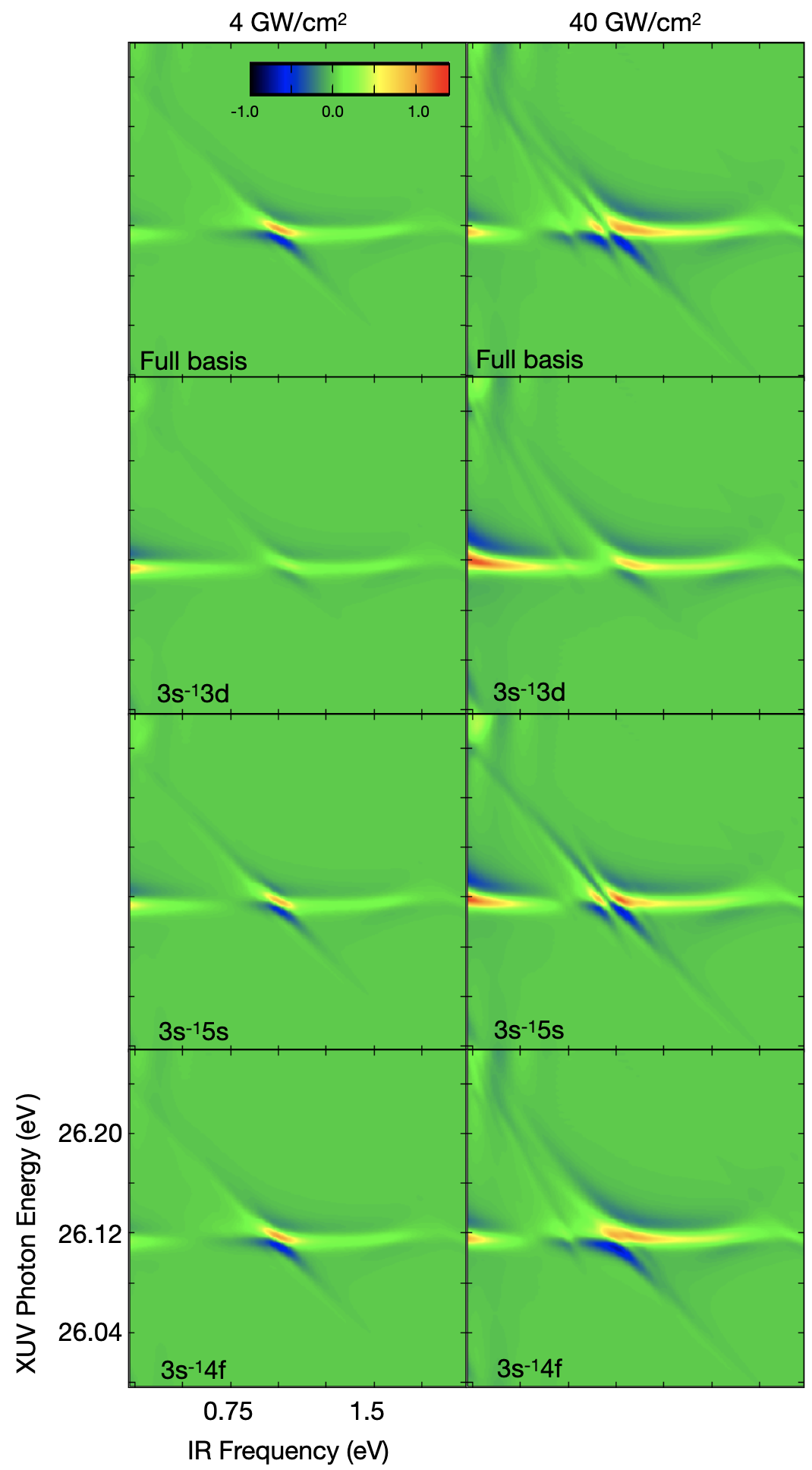}
    \caption{Normalized transient absorption signal, represented as $\Delta\sigma \,I_0/I_{\textsc{IR}}$ with $I_0=$40\,GW/cm$^2$, as a function of the dressing laser frequency at a fixed zero time delay. The signal is computed \emph{ab initio} for two different peak intensities of the dressing laser field: 4\,GW/cm$^2$, 40\,GW/cm$^2$. The figure is organized to demonstrate the impact of different basis configurations on the absorption signal. The top row presents results computed using the full basis. Subsequent rows illustrate the effect of selectively omitting certain autoionizing states from the basis- specifically, the $3s^{-1}3d$, $3s^{-1}5s$, and $3s^{-1}4f$ states.}
    \label{fig:FDepResCont}
\end{figure}
To determine which transitions contribute to the observed spectral features, we selectively exclude specific autoionizing states from the ES-shifted basis used in the simulation. Figure~\ref{fig:FDepResCont} demonstrates this method by presenting a normalized transient absorption signal, $\Delta\sigma$, defined as $\Delta\sigma=\Delta\sigma(\omega_{\textsc{xuv}};\omega_{\textsc{ir}},I_{\textsc{ir}},\tau)\equiv\sigma(\omega_{\textsc{xuv}};\omega_{\textsc{ir}},I_{\textsc{ir}},\tau)-\sigma(\omega_{\textsc{xuv}};\omega_{\textsc{ir}},I_{\textsc{ir}},-\infty)$. This signal is obtained using either the full basis or by omitting one of the AIS that lead to a relevant LIS, at two different intensities of the dressing laser, $I_{\textsc{IR}}=4$\,GW/cm$^2$ and $I_{\textsc{IR}}=40$\,GW/cm$^2$. 
Since $\Delta\sigma$ scales linearly with the laser intensity to lowest order, the $\Delta\sigma$ for $I_{\textsc{IR}}=4$\,GW/cm$^2$ has been scaled by a factor of 10 for easier comparison with the spectrum computed at the higher intensity. At the lowest intensity, LIS associated with two-photon transitions are barely discernible. Indeed, removing the $3s^{-1}4f$ state from the simulation at $I_{\textsc{IR}}=4$\,GW/cm$^2$ has a negligible effect on the spectrum. We also notice that the stronger change in the spectrum is due to the $3s^{-1}3d$ state. At higher intensities, the spectrum develops features associated to two new LIS. As suggested by Fig.~\ref{fig:FreqSweep}, the feature intersecting the avoided crossing between the $3s^{-1}4p$, $3s^{-1}5s$, and $3s^{-1}3d$ states is likely due to the $3s^{-1}4f$ LIS. Indeed, by eliminating this AIS from the basis, that feature disappears. A similar calculation enables us to attribute the second new feature, intersecting the $3s^{-1}4p$ line at approximately $\omega_{\textsc{IR}}\simeq 0.75$\,eV, to the $3s^{-1}5p$ AIS.
A common feature to the spectra shown in Figs.~\ref{AbES}-\ref{fig:FreqSweep} is that the AIPs profiles overlap, interfering constructively or destructively with each other. Across the next two sections we will show how extending the Jaynes-Cummings model to autoionizing states leads to a functional expression for overlapping resonances that can be used to fit the experimental measurement and which was in fact essential to quantitatively compare the parameters of the $3s^{-1}4p$\,---\,$3s^{-1}3d$ AIPs in Argon with those predicted by our \emph{ab initio} simulations~\cite{Harkema2022}.

\section{\label{sec:JCbasic} Simplified model for radiatively coupled autoionizing states}

In this section, we derive the expression for the energy and width of two autoionizing states coupled by a moderately intense cw dressing IR laser, under the premise that the AIPs resulting from the coupling are isolated, i.e., separated by an energy gap much larger than their decay width.  To monitor the formation of these AIPs, we evaluate the XUV absorption cross section of the dressed system, from the ground state, in which these AIPs are expected to leave a distinct optical signature. For simplicity, we will neglect the influence of the IR dressing field on the ground state, which is a fair assumption if the excitation energy of the ground state is much larger than the IR central frequency. We will also assume that direct ionization from the ground state to the continuum is negligible. The more general case for overlapping resonances in the presence of a background ionization amplitude is examined in Sec.~\ref{sec:JCAdvanced}.
\begin{figure}[hbtp!]
\centering
\includegraphics[width=0.7\columnwidth]{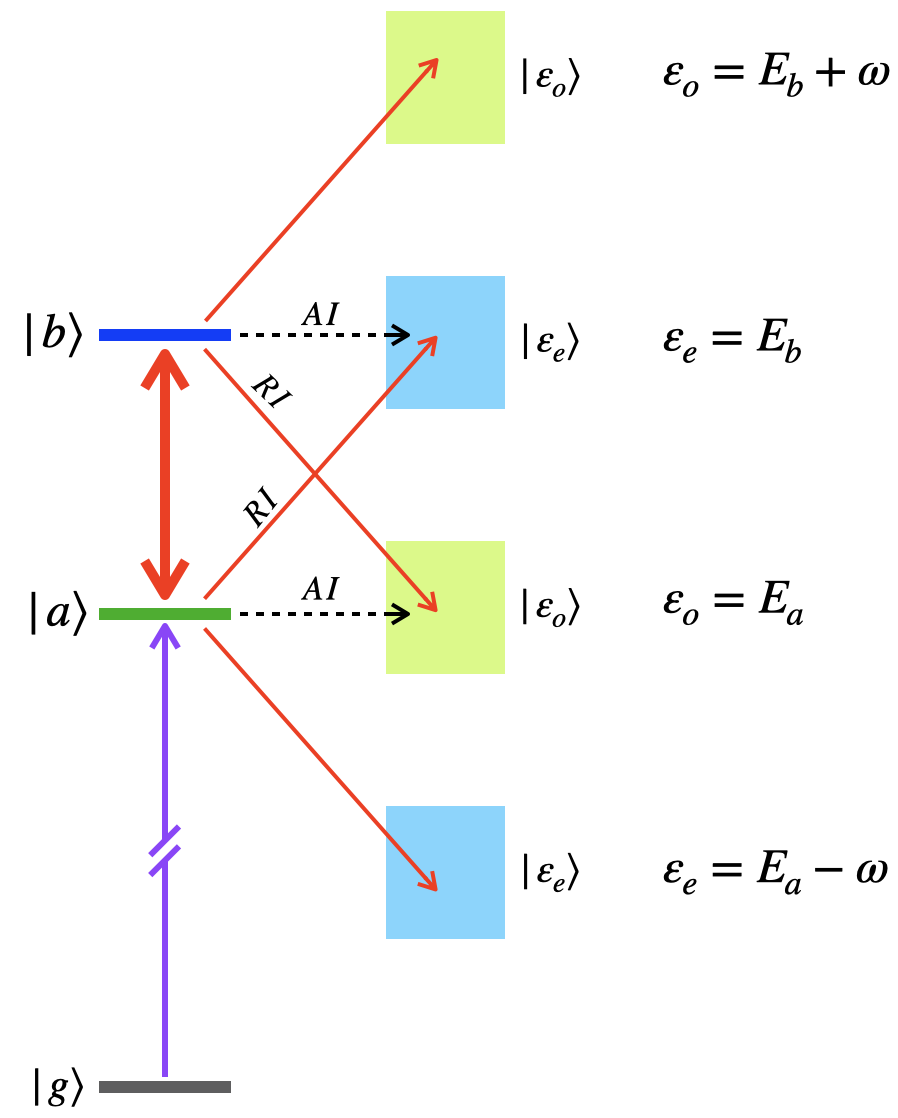}
\caption{\label{fig:LevelI}\footnotesize Diagram depicting the various radiative and nonradiative couplings in our model system, which includes a ground state, two AIS of opposite parity, and their respective continua. The radiative coupling from the AIS to the continuum, as well as between the ground and the bright AIS, is treated perturbatively, while the coupling between the localized component of the field-free autoionizing resonances that form the autoionizing polaritons is treated nonperturbatively.
}
\end{figure}
To model these assumptions, we start partitioning the field-free Hamiltonian into a reference Hamiltonian $\hat{H}_0$, and a configuration-interaction term $\hat{V}$, $\hat{H}=\hat{H}_0 + \hat{V}$, both of which are Hermitian. The reference Hamiltonian $\hat{H}_0$ is chosen to feature three bound and two sets of continuum eigenstates. The bound states consist in the ground state, $\hat{H}_0|g\rangle = |g\rangle E_g$, with even parity, and two autoionizing states, $\hat{H}_0|a\rangle = |a\rangle E_a$ and $\hat{H}_0|b\rangle = |b\rangle E_b$, with odd and even parity, respectively. The two sets of continuum eigenstates, $\hat{H}_0 | \varepsilon_e\rangle =| \varepsilon_e\rangle \varepsilon_e$ with even parity and $\hat{H}_0 | \varepsilon_o\rangle =| \varepsilon_o\rangle \varepsilon_o$ with odd parity, are defined for energies above a given threshold $E_{\mathrm{th}}$, $\varepsilon_{e/o}>E_{\mathrm{th}}$. In the spectral basis of $\hat{H}_0$, $\hat{V}$ is assumed to only have the following non-zero matrix elements are $V_{a\varepsilon_o}=\langle a | \hat{V} | \varepsilon_o\rangle=V_{\varepsilon_oa}^*$ and $V_{b\varepsilon_e}=\langle b | \hat{V} | \varepsilon_e \rangle=V_{\varepsilon_eb}^*$. Both $E_a$ and $E_b$ are set above $E_{\mathrm{th}}$, leading to the Auger decay of these two states through their coupling to the continuum.
As an initial approximation, the Auger decay rates can be computed using Fermi's golden rule~\cite{Sakurai}
\begin{equation}
\Gamma_{a} = 2\pi \left|\langle {\varepsilon}_o | \hat{V} | a \rangle\right|^2,\quad
\Gamma_{b} = 2\pi \left|\langle {\varepsilon}_e | \hat{V} | b \rangle\right|^2.
\end{equation}
 In the numerical examples examined in this section, $E_g$, $E_a$, and $E_b$ match the energies of the ground state and of the $3s^{-1}4p$ and $3s^{-1}3d$ states of argon, respectively~\cite{Harkema2022,Yanez-Pagans2022}.
For weak XUV pulses, as it is normally the case in attosecond transient absorption experiments, the XUV photoionization amplitude from the ground state to the continuum can be evaluated using first-order perturbation theory. As a result, the observed optical density is directly proportional to the square modulus of the dipole transition amplitude from the ground state to the resonant continuum. In the absence of dressing radiation and assuming that the direct radiative amplitude from the ground state to the reference continuum $|\varepsilon_o\rangle$ is negligible, therefore, the XUV spectrum of the system will exhibits a single bright Lorentzian resonance peak centered near $E_a$, with width $\Gamma_a$. 

In the presence of the IR field, the two states $|a\rangle$ and $|b\rangle$ are radiatively coupled with each other, as well as to the ionization continuum with opposite parity. Figure~\ref{fig:LevelI} schematically represents these energy levels and their radiative couplings. The interplay between Autler-Townes splitting, autoionization decay, and radiative decay will affect the position and width of the peaks visible in the XUV absorption spectrum. To better highlight the effect of the IR field on the AIPs widths, in this section we will assign identical Auger decay rates and radiative coupling strengths to the continuum for these two autoionizing states. 
To describe the strong radiative coupling between states $|a\rangle$ and $|b\rangle$ due to the IR field, and the effect of the coherence between radiative and autoionization transitions on the decay of the resulting laser-dressed states, it is convenient to use a quantized formalism for the radiation rather than a semiclassical description of the IR field. For large number of photons and for laser fields described by coherent states, the two approaches are equivalent~\cite{Cohen-Tannoudji1996}. 
The vector-potential operator for a single IR radiation mode with frequency $\omega$ and polarization $\hat{\epsilon}$ (the hat, in this case, indicates a unit vector rather than an operator), in a cubic box with edge length $L$ and periodic boundary conditions, is defined as~\cite{CohenTannoudjiAtomsAndPhotons}
\begin{equation}
\hat{\vec{A}}_{\textsc{IR}}=\left(\frac{2\pi c}{L^3 k}\right)^{1/2}
\left(
\hat{\epsilon}\,\hat{\eta}+
\hat{\epsilon}^*\,\hat{\eta}^\dagger
\right),
\end{equation}
where $k=\omega/c$, $c$ is the speed of light in vacuum, and $\hat{\eta}^\dagger$ and $\hat{\eta}$ are the photon creator and annihilation operators, respectively, with $\hat{\eta}|n\rangle = \sqrt{n} |n-1\rangle$ and $\hat{\eta}^\dagger|n\rangle = \sqrt{n+1} |n+1\rangle$.  The free-radiation Hamiltonian for this mode is $\hat{H}_R = \omega \hat{\eta}^\dagger \hat{\eta}$. A photon-state $|n\rangle$ thus has energy $n\omega$, with $\hat{H}_R|n\rangle= n\omega|n\rangle$. The average number of photons per unit volume, $n/L^3$, can be determined from the laser intensity $I$ as $n/L^3 = I/(\omega c)$. The minimal-coupling radiation-matter interaction Hamiltonian is given by $\hat{H}_I= \vec{A}_{IR}\cdot\hat{\vec{p}} / c$, $\hat{\vec{p}}$ is the electron momentum, and we assume the dipole approximation ($\omega r/c\ll1$). 

For a moderately intense dressing laser (e.g., $I\sim10^{11}$\,W/cm$^2$ at $\lambda\sim 1\,\mu$m ) the radiative coupling of either $|a\rangle$ or $|b\rangle$ to the continuum can become weak enough to be described perturbatively. Furthermore, the ponderomotive energy is a negligible fraction of the laser frequency, we can safely disregard the laser dressing of the continuum, including radiative continuum-continuum transitions. Nevertheless, for sufficiently small detuning $\delta = E_b-E_a-\hbar\omega$ from the resonant transition between $|a\rangle$ and $|b\rangle$, these two states would become strongly radiatively coupled.
To limit the interaction to these processes, we modify the interaction Hamiltonian $\hat{H}_I$ to an \emph{ad hoc} operator $\hat{\bar{H}}_I$ that deliberately excludes laser dressing of the ground state and the continuum,
\begin{equation}
\begin{split}
\hat{\bar{H}}_I 
&= \hat{H}_I - \hat{P}_g \hat{H}_I - \hat{H}_I \hat{P}_g - \hat{Q} \hat{H}_I \hat{Q},
\end{split}
\end{equation}
where $\hat{P}_g=|g\rangle \langle g|$, and 
$\hat{Q}=\int d\varepsilon\left( |\varepsilon_{e}\rangle \langle \varepsilon_{e}|+|\varepsilon_{o}\rangle \langle \varepsilon_{o}|\right)$ are the projectors on the ground and the reference continuum states, respectively. In summary, the total Hamiltonian for matter and the dressing field is
\begin{equation}
\hat{H} = \hat{H}_0 + \hat{V} + \hat{H}_R + \hat{\bar{H}}_I,
\end{equation}
and it acts on the tensor-product space $\mathcal{H}=\mathcal{H}_{el}\otimes\mathcal{H}_r$, where $\mathcal{H}_{el}=\mathrm{Span}\{|g\rangle, |a\rangle, |b\rangle, |{\varepsilon}_e\rangle, |{\varepsilon}_o\rangle\}_{\varepsilon_{e/o}\in [E_{\mathrm{thr}},\infty]}$ and $\mathcal{H}_R=\mathrm{Span}\{|n\rangle,\,n\in\mathbb{N}\}$ are the Hilbert spaces for matter and radiation, respectively. 
It is useful to decompose $\hat{H}$ into a component $\hat{H}_0'$ that accounts for the non-perturbative radiative coupling between $|a\rangle$ and $|b\rangle$, and a residual perturbation $\hat{V}'$, responsible for the radiative and non-radiative coupling of these states to the continuum, 
\begin{equation}\label{eq:HPartitioning}
\begin{split}
\hat{H} &= \hat{H}_0' + \hat{V}'\\
\hat{H}_0' &= \hat{H}_0 + \hat{H}_R + \hat{P}\hat{H}_I\hat{P},\\
\hat{V}' &= \hat{V} + \hat{P} \hat{H}_I \hat{Q}  + \hat{Q}\hat{H}_I\hat{P},
\end{split}
\end{equation}
where $\hat{P}=|a\rangle \langle a| + |b\rangle \langle b|$.
In this context, the continuum states $|{\varepsilon}_{o/e},n\rangle = |{\varepsilon}_{o/e}\rangle\otimes|n\rangle$ are already eigenstates of $\hat{H}_0'$. The other eigenstates are linear combinations of the states $|a,n\rangle \equiv |a\rangle \otimes |n\rangle$ and $|b,n'\rangle$, which are entangled states of matter and radiation, commonly known as polaritons~\cite{FoxSolids}. To further simplify, we assume that only configurations of the form $|a,n\rangle$ and $|b,n-1\rangle$ significantly mix, effectively leading us to the well-known Jaynes-Cummings model \cite{Jaynes1963}. This assumption is equivalent to the rotating-wave approximation in the semi-classical description of the Rabi oscillations~\cite{Rabi1937}.

The eigenstates of $H_0'$ are obtained by diagonalizing the Hamiltonian in the $2\times 2$ basis of $|a,n\rangle$ and $|b,n-1\rangle$,
\begin{equation}
\begin{split}
\langle a,n|\hat{H}|a,n\rangle&=E_a+n\omega\\
\langle b,n-1|\hat{H}|b,n-1\rangle&=E_b+(n-1)\omega\phantom{\sqrt{\frac{2\pi}{ c}}}\\
\langle a,n|\hat{H}|b,n-1\rangle&=\sqrt{\frac{2\pi}{ c}}\,\frac{I^{1/2}}{\omega}
p_{a b},
\end{split}
\end{equation}
where $p_{a b} =\langle a|\hat{\epsilon}\cdot\hat{\vec{p}}|b\rangle$. 
The two eigenvectors of $H_0'$,
\begin{equation}
|\Psi^\pm\rangle = |a,n\rangle\, C_a^\pm + |b,n-1\rangle\, C_b^\pm,     
\end{equation}
with eigenvalues $E_\pm = E_a + n\omega + \varepsilon_\pm$, 
satisfy the following secular equation,
\begin{equation}
\left(
\begin{array}{cc}
0  & \gamma_{ab}   \\
\gamma_{ab}^* & \delta
\end{array}
\right)
\left(
\begin{array}{c}
C_a^\pm  \\
C_b^\pm
\end{array}
\right) = \varepsilon_\pm \left(
\begin{array}{c}
C_a^\pm  \\
C_b^\pm
\end{array}
\right),
\end{equation}
where $\gamma_{ab} = \sqrt{\frac{2\pi I}{\omega^2c}}\,p_{a b}$, $\delta=E_b-E_a-\omega$, and
\begin{equation}\label{eq:Epm}
    \varepsilon_{\pm} = \frac{\delta}{2} \pm \frac{\Omega}{2},\qquad {\Omega} = \sqrt{\delta^2+\Omega_0^2},\quad \Omega_0 = 2\gamma_{ab},
\end{equation}
with ${\Omega}$ being the familiar Rabi frequency. 
The corresponding eigenvectors, therefore, are
\begin{equation}\label{eq:Psipm}
\begin{split}
    |\Psi^+\rangle &= \phantom{-}\cos\theta \,|a;n\rangle + \sin\theta \,|b;n-1\rangle, \\
    |\Psi^-\rangle &= -\sin\theta \,|a;n\rangle + \cos\theta \,|b;n-1\rangle,
\end{split}
\end{equation}
where $\theta = \arctan[\Omega_0/(\Omega-\delta)]$. 

\begin{figure}[hbtp!]
\centering
\includegraphics[width=0.8\columnwidth]{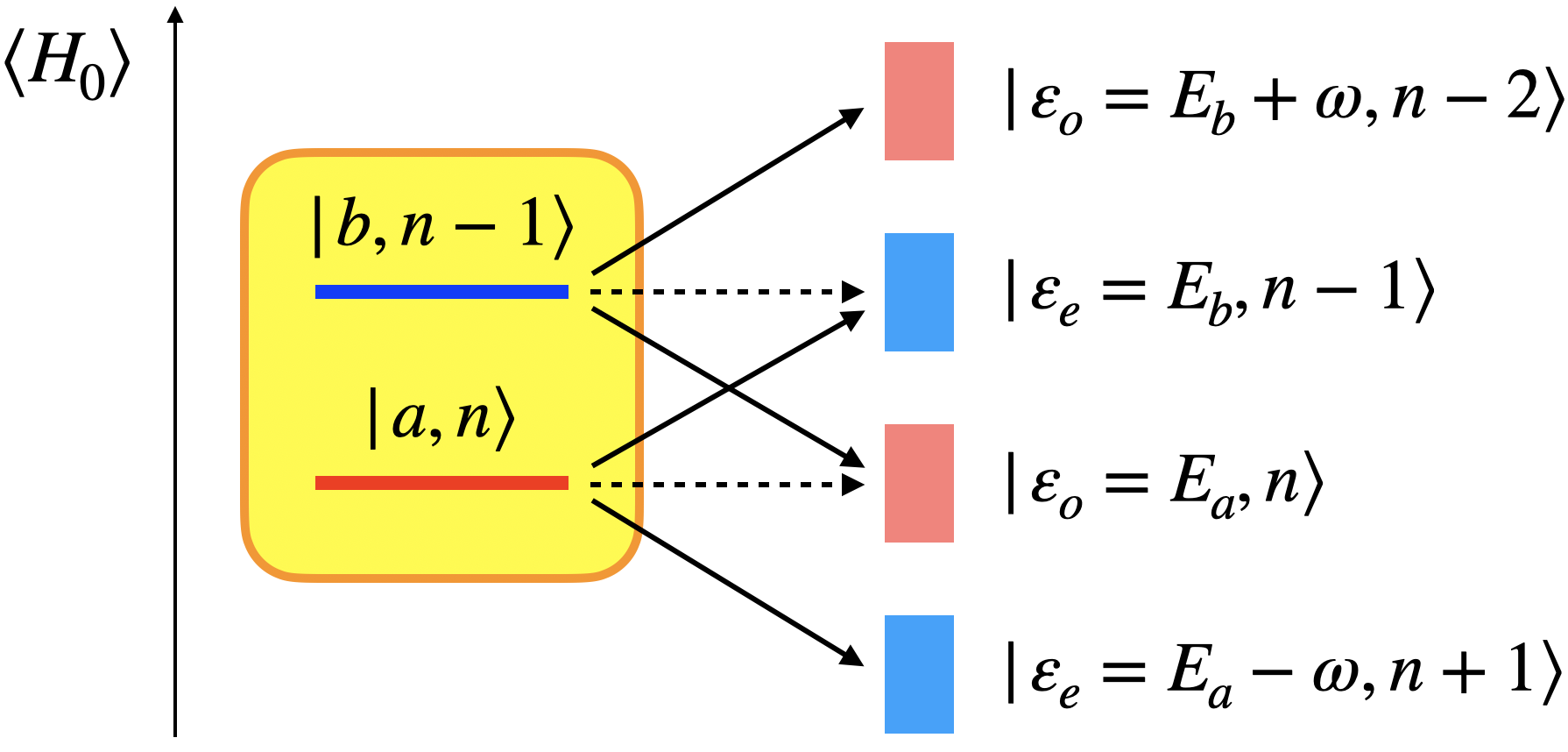}
\caption{\label{fig:PolaritonDecayPaths}\footnotesize Decay paths of one autoionizing polariton (yellow box on the left). The vertical axis represents the expectation value of the non-interacting electronic Hamiltonian. The energies of the two components of the polariton are shown separately for clarity. The four possible final continuum channels to which the polariton decays are shown on the right. Dashed arrows represent Auger decay, where the number of photons does not change, while continuum arrows represent either the stimulated emission or absorption of one photon that accompanies the ionization of a polaritonic component, for a total of six transition paths. For each of the two central final continuum channels, two distinct paths interfere: Auger decay of one component and the photoionization (in either absorption or emission) of the other component, leading the same final energy and parity in the continuum.}
\end{figure}

Figure~\ref{fig:PolaritonDecayPaths} illustrates the six different paths for the decay of a polaritonic state. If the two polaritonic states are energetically well separated, we can estimate their decay rates using Fermi's golden rule, 
\begin{equation}
\begin{split}
\Gamma^{\pm} &= \sum_{\pi=e,\,o}\sum_{n'} \Gamma^{\pm}_{\varepsilon_\pi,n'},\\
\Gamma^{\pm}_{\varepsilon_\pi,n'} &= 2\pi \left|\langle \varepsilon_\pi, n' | \hat{V}' | \Psi_{\pm} \rangle \right|^2.
\end{split}
\end{equation}
In this context, the summation is limited to four possible final channels, differentiated by the parity of the matter continuum, and by the number of IR photons,
\begin{equation}
\begin{split}
\Gamma^\pm &= \Gamma^\pm_{\varepsilon_e=E_b,n-1}+\Gamma^\pm_{\varepsilon_o=E_a,n}+\\
&+\Gamma^\pm_{\varepsilon_e=E_a-\omega,n+1}+\Gamma^\pm_{\varepsilon_o=E_b+\omega,n-2}.
\end{split}
\end{equation}
Two of these decay channels, $|\Psi^\pm\rangle \to |\varepsilon_o=E_a-\omega,n+1\rangle$ and $|\Psi^\pm\rangle \to|\varepsilon_e=E_b+\omega,n-2\rangle$, result from individual quantum paths involving either the coherent emission or absorption of a photon. As these paths do not interfere with any others, their contribution adds incoherently to the total rate.
In the remaining two channels, on the other hand, the Auger decay amplitude from each polaritonic component ($|a,n\rangle$ or $|b,n-1\rangle$) interferes with one of the radiative ionization amplitudes from the other component, thus enhancing or suppressing the overall decay rate.
The partial decay rates for the two channels where autoionization and radiative ionization interfere are expressed as follows:
\begin{equation}
\Gamma^{\pm}_{\varepsilon_e=E_b,n-1} 
=2\pi \left|\, \frac{1}{\omega}\, \sqrt{\frac{2\pi I}{c}}\,p_{\varepsilon_e a}\, C_a^\pm + V_{\varepsilon_e,b}\,C_b^\pm\,\right|^2
\end{equation}
where we have used the relation
\begin{equation}
    \langle \varepsilon_e,n-1|\hat{V}'|a,n\rangle =\frac{1}{\omega} \sqrt{\frac{2\pi I}{c}}\,p_{\varepsilon_e a}.
\end{equation}
Recognizing that $\Gamma_{AI,b} = 2\pi |V_{\varepsilon_e,b}|^2$ is the Auger width of the $|b\rangle$ state
and introducing $\Gamma_{PI,a} = 4\pi^2 I p^2_{E_\beta,a}/\omega^2$ as the decay rate associated to the photoionization of $|a\rangle$, the width $\Gamma_{\varepsilon_e=E_b,n-1}^\pm$ can be reformulated as
\begin{equation}\label{eq:WidthInterference}
\begin{split}
\Gamma^{\pm}_{\varepsilon_e=E_b,n-1} &=\left|\,\Gamma^{1/2}_{PI,a}\, C_a^\pm+\Gamma^{1/2}_{AI,b}\, C_b^\pm\,\right|^2=\\
&=\Gamma_{PI,a}\left|C_a^\pm\right|^2 + \Gamma_{AI,b}\left|C_b^\pm\right|^2 +\\ 
&+ 
2\,\Gamma^{1/2}_{PI,a}\,\Gamma^{1/2}_{AI,b}\,\,\Re\left( {C_a^\pm}^*\,C_b^\pm\right).
\end{split}
\end{equation}
Similarly, for the other channel exhibiting interference,
\begin{equation}
\Gamma^{\pm}_{\varepsilon_o=E_a,n} =\left|\Gamma^{1/2}_{AI,a}\, C_a^\pm+\Gamma^{1/2}_{PI,b}\, C_b^\pm\right|^2.
\end{equation}
The two remaining channels do not interference with any other, 
\begin{equation}
\begin{split}
\Gamma^{\pm}_{b,E_\alpha-\omega,n+1} &=\Gamma_{PI,a} \left|C_a^\pm\right|^2,\\
\Gamma^{\pm}_{a,E_\beta+\omega,n-2} &=\Gamma_{PI,b} \left|C_b^\pm\right|^2.
\end{split}
\end{equation}
The total ionization rate is thus given by
\begin{equation}\label{JCMIonRate}
\begin{split}
\Gamma^\pm =& \,(2\Gamma_{PI,a}+\Gamma_{AI,a})\left|C_a^\pm\right|^2 +\\ +&\,(2\Gamma_{PI,b}+\,\Gamma_{AI,b})\left|C_b^\pm\right|^2 + \\
+&\,2\,(\Gamma^{1/2}_{PI,a}\Gamma^{1/2}_{AI,b}+\Gamma^{1/2}_{AI,a}\Gamma^{1/2}_{PI,b})\,\,\,\Re\left( {C_a^\pm}^*C_b^\pm\right).
\end{split}
\end{equation}
Depending on the detuning $\delta$ and laser intensity $I$, the interference term can either enhance or suppress the decay rate to the channel. This is the mechanism with which one of the two polaritonic branches can be stabilized. Notice that, at sufficiently low laser intensities, the interference term becomes dominant over the quadratic photoionization term. 

\begin{figure}
    \includegraphics[width=\columnwidth]{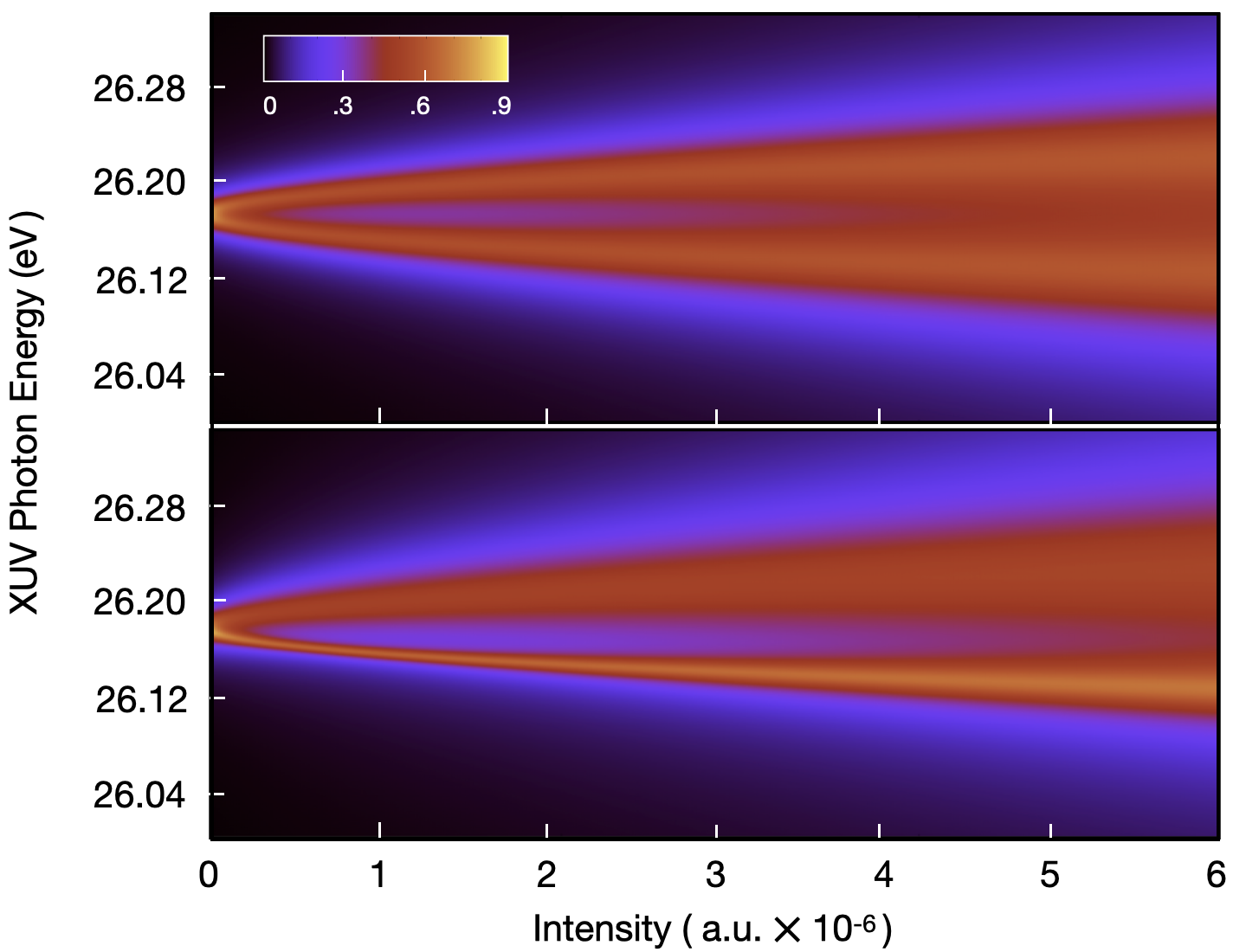}
    \caption{Simulated absorption spectrum (in arbitrary units) as a function of the dressing laser intensity. This model neglects the radiative ionization of the ground state directly to the continuum, therefore the AIPs appear as a pair of Lorentzian functions centered at $\varepsilon_\pm$, taken from~\eqref{eq:Epm}, and have widths $\Gamma^\pm$ given by~\eqref{JCMIonRate}. In the upper plot equation~\eqref{JCMIonRate} has been modified such that the AI and PI terms summed incoherently.
    \label{fig:LorentzianAIPs}}
\end{figure}
\begin{figure}
    \centering
    \includegraphics[width=\columnwidth]{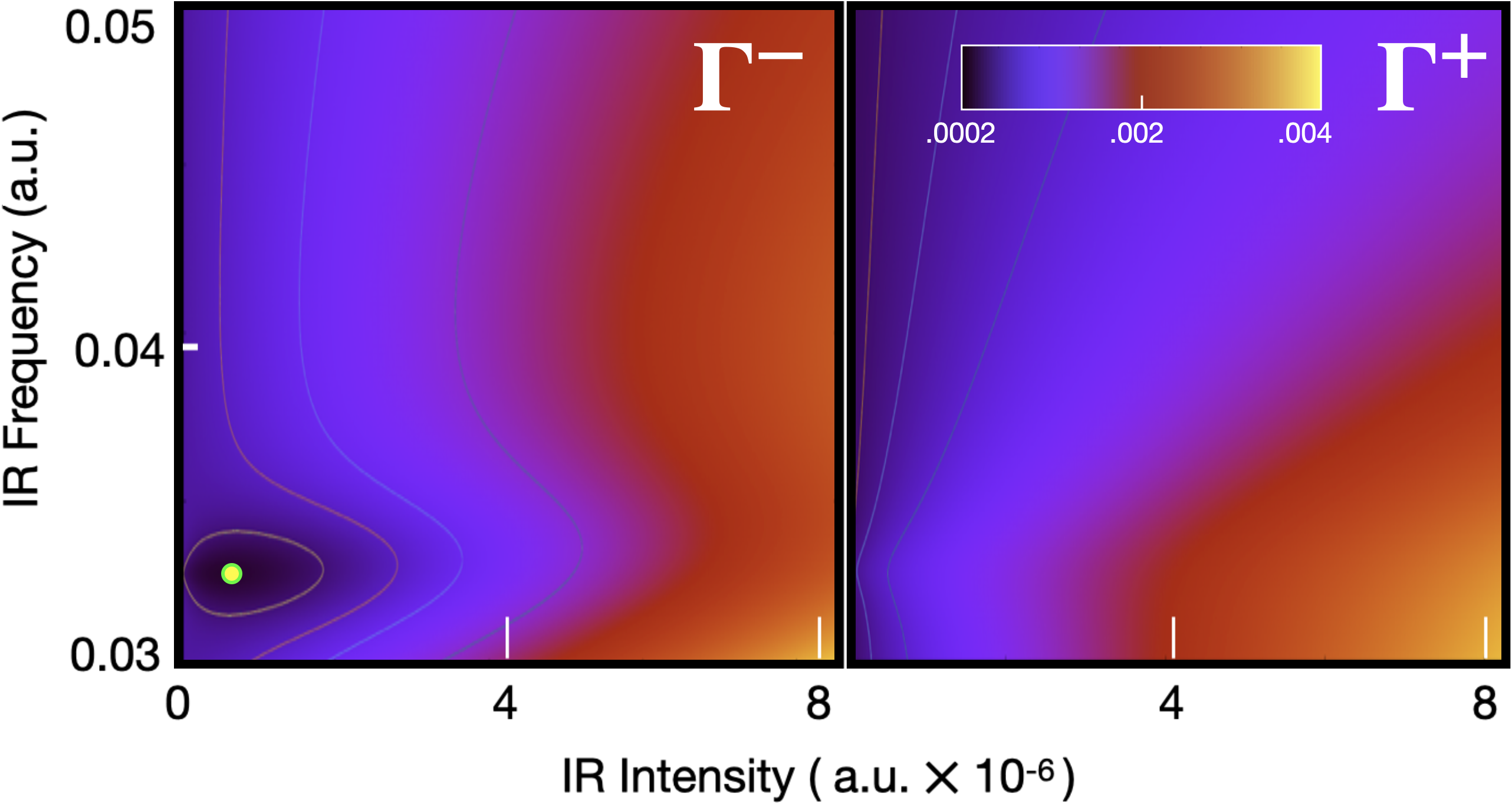}
    \caption{Width of the lower (a) and upper (b) AIP branch as a function of laser detuning and intensity. In the lower branch, a minimum in the width occurs around $\delta=0$, at a critical laser intensity value of $7\times10^{-6}\,$a.u. (yellow circle). \label{fig:GammaMinusIvF}
}
\end{figure}

Figure~\ref{fig:LorentzianAIPs} shows a simulated absorption spectrum $\sigma(E;I)$, as a function of the laser intensity. The spectrum is computed using the following equation:
\begin{equation}
\begin{split}
\sigma(E;I) = \frac{|p_{ag}|^2}{2\pi}\Bigg[&\,\,\frac{|C_a^+|^2\,\Gamma^+}{(E-E_+)^2+(\Gamma^+/2)^2}\,\,+\\
+&\,\,\frac{|C_a^-|^2\,\Gamma^-}{(E-E_-)^2+(\Gamma^-/2)^2}\,\,\Bigg],
\end{split}
\end{equation}
where $p_{ag}=\langle a |p|g\rangle$. As detailed at the beginning, the present model assumes equal Auger widths for states $|a\rangle$ and $|b\rangle$ and neglects direct ionization from the ground to the continuum, leading to Lorentzian profiles for the AIP branches. At higher laser intensities, the two polaritonic branches become clearly distinguishable. 
The upper plot excludes the interference term from equation~\eqref{JCMIonRate}, causing the radiative and nonradiative terms to add incoherently, resulting in equal spectral widths for the upper and lower AIP branches. In contrast, the lower plot includes the interference term, making the upper polaritonic branch visibly broader than the lower branch.
This demonstrates that the modulation in the widths of the polaritonic branches is due to interference between competing ionization pathways. As equation~\eqref{JCMIonRate} suggests, for critical values of the detuning and laser intensity, the width of one of the AIP branches can be minimized, thus stabilizing one of the AIP against ionization. This phenomenon is illustrated in Figure~\ref{fig:GammaMinusIvF}, which shows the width of the lower and upper AIP as a function of IR field frequency and intensity. The width of the lower branch reaches a minimum when the undressed states $|a\rangle$ and $|b\rangle$ are in resonance and the laser intensity realizes the destructive-interference condition in Eq.~\eqref{eq:WidthInterference}. This finding, which gives a qualitative explanation of the stabilization observed in \emph{ab initio} simulations described in Sec.~\ref{sec:abinitio} and in the experiment (see Fig.\,2d,e in~\cite{Harkema2022}), is the main result of this section.

\section{Model for Overlapping AIPs}\label{sec:JCAdvanced}
In the model derived in the previous section, we neglected the direct ionization to the continuum and did not consistently describe the interference terms that should be observed when the two AIP overlap. While it reproduced the AIP stabilization effect, therefore, this simplified model cannot correctly reproduce the spectral profile of the AIPs observed in a real experiment. This shortcoming is particularly pronounced when trying to explain AIPs below the $3s^{-1}$ threshold in argon, where the most prominent autoionizing states appear as window resonances rather than as Lorentzian's profiles. In this section, we expand our model by considering the interference between direct photoionization amplitudes and resonant-ionization amplitudes. This extension enables our model to reproduce Fano-like profiles of isolated resonances, as well as the more complex interference profile of two overlapping polaritons. This extension, in particular, is essential to reproduce window resonances, the accurate spectral profile of the two polaritons when the Rabi frequency is comparable to or smaller than the polaritons' widths, and the case of polaritons emerging from the strong radiative coupling of multiple autoionizing states. 

As in the previous section, we start with the partitioning of the Hamiltonian along the lines of \eqref{eq:HPartitioning}. We assume that the ionization of the laser-dressed target can be described in terms of a finite number of localized matter-radiation states, $( |a;n\rangle, |b;n-1\rangle, ... )$, which give rise to autoionizing resonances upon interaction with the continuum. Additionally, we consider a finite number of electronic-continuum channels $|\boldsymbol{\varepsilon}\rangle = ( |\alpha\varepsilon\rangle, |\beta\varepsilon\rangle, ... )$, where $\varepsilon$ represents the total energy (photoelectron energy plus radiation field energy) and the channel labels $\alpha$ include discrete quantum numbers identifying the state of the ion, the photoelectron's angular distribution, their angular and spin coupling, as well as the number of photons in the dressing radiation. 

As before, we continue assuming that the ground state $|g\rangle$ is not coupled by the dressing field to either the localized or the continuum states. Therefore, the initial state can be represented as $|g,n\rangle=|g\rangle\otimes|n\rangle$ for a suitable choice of $n$, presumed to be sufficiently large. We will consider only perturbative transitions from the ground state to the continuum, induced by the absorption of a single XUV photon.
The Hamiltonian's matrix elements within the bound and the continuum sector of the matter states are then given by
\begin{equation}
\langle a,n|\hat{H}_0'| a',n'\rangle = (E_a+n\omega)\,\delta_{aa'}\delta_{nn'}+\sqrt{\frac{2\pi I}{c\omega^2}}p_{a a'},
\end{equation}
\begin{equation}
\langle\alpha\varepsilon|\hat{H}_0'|\beta\varepsilon'\rangle = \varepsilon\,\delta_{\alpha\beta}\delta(\varepsilon-\varepsilon')+\langle\alpha\varepsilon|H_I|\beta\varepsilon'\rangle,
\end{equation}
whereas the interaction between bound and continuum states is
\begin{equation}\label{eq:RadiativeCoupling}
\langle a,n|\hat{V}+\hat{H}_I|\alpha\varepsilon\rangle =\sqrt{\frac{2\pi I}{c\omega^2}}p_{a \varepsilon_\alpha}\delta_{|n-n_\alpha|,1},
\end{equation}
where $\varepsilon_\alpha=\varepsilon-n_\alpha\omega$.

To determine the eigenstates of the full Hamiltonian, we apply Fano's formalism for multiple bound states interacting with multiple continua~\cite{Fano1961}. Additionally, we introduce a simplifying assumption: for the intensities of the dressing IR laser considered, the continuum-continuum radiative transitions can be neglected. 

Indeed, at sufficiently moderate intensities, the rate of continuum-continuum radiative transitions is negligible compared to the Auger decay rate. Prior to applying Fano's formalism, it is convenient to diagonalize the bound sector of the Hamiltonian through an orthogonal transformation. We define a set of states $|\Phi_i\rangle$ such that
\begin{eqnarray}
&&|\Phi_i\rangle = \sum_{a,n} |a,n\rangle O_{a,n;i}\\
&&\langle \Phi_i | \hat{H}_0' |\Phi_j\rangle = E_i \delta_{ij},\qquad
\langle \Phi_i | \Phi_j\rangle = \delta_{ij}.
\end{eqnarray}

Our goal is to find expressions for the stationary wave functions in the continuum, with total energy E, that match prescribed boundary conditions specified by a continuum-channel label $\alpha$. We represent these wave functions as:
\begin{equation}\label{eq:PsiGeneral}
    |\Psi_{\alpha E}\rangle = |\mathbf{\Phi}\rangle\mathbf{f}_{\alpha E}  + \ 
     \sum_\gamma\int d\varepsilon | \gamma\varepsilon \rangle c_{\gamma\varepsilon,\alpha E},
\end{equation}
 where $|\mathbf{\Phi}\rangle = (|\Phi_1\rangle,\,|\Phi_2\rangle,\ldots)$, and $\mathbf{f}_{\alpha E}$ and $c_{\gamma\varepsilon,\alpha E}$ are expansion coefficients. To determine these coefficients, we impose the condition that $|\Psi_{\alpha E}\rangle$ satisfies the secular equation for the total Hamiltonian,
\begin{equation}
    (E-\hat{H})|\Psi_{\alpha E}\rangle = 0.
\end{equation}
This equation for the wave function is converted into a system of equations for its expansion coefficients by projecting it on the basis of localized and continuum states. The projection on the bound states reads
\begin{equation}\label{Orig1}
(E-\mathbf{E})\mathbf{f}_{\alpha E} - \sum_\gamma\int d\varepsilon \langle \mathbf{\Phi}|\hat{V}'|\gamma\varepsilon\rangle \,c_{\gamma\varepsilon,\alpha E} = 0,
\end{equation}
where $\mathbf{E}_{ij}=\delta_{ij}E_i$, 
using the relations $\langle\boldsymbol{\Phi}|\hat{H}|\boldsymbol{\Phi}\rangle=\boldsymbol{E}$, and $\langle\boldsymbol{\Phi}|\gamma\varepsilon\rangle=\boldsymbol{0}$.
The projection on the continua is given by
\begin{equation}\nonumber
    -\langle\beta \varepsilon|\hat{V}'|\mathbf{\Phi}\rangle\mathbf{f}_{\alpha E} + \sum_\gamma\int d\varepsilon' \langle \beta\varepsilon|E-\hat{H}|\gamma\varepsilon'\rangle\,c_{\gamma\varepsilon',\alpha E} = 0,
\end{equation}
which can be reformulated as:
\begin{equation}\label{Orig2}
    (E-\varepsilon)\,c_{\beta\varepsilon,\alpha E} = \langle \beta \varepsilon | \hat{V}' |\mathbf{\Phi}\rangle \mathbf{f}_{\alpha E}.
\end{equation}
From Eq.~\eqref{Orig2}, it is possible to express the expansion coefficients on the continuum states in terms of the coefficients of the autoionizing states. The general solution to~\eqref{Orig2} is given by a particular solution $c'_{\beta\varepsilon,\alpha E}$ of that equation, e.g., in principal part,
\begin{equation}\nonumber
c'_{\beta\varepsilon,\alpha E} =\frac{\mathcal{P}\langle \beta \varepsilon | \hat{V}' |\mathbf{\Phi}\rangle \mathbf{f}_{\alpha E}}{E-\varepsilon},
\end{equation}
plus a solution to the associated homogeneous equation $(E-\varepsilon)\,c^{\mathrm{Hom}}_{\beta\varepsilon,\alpha E} =0$, e.g., $c^{\mathrm{Hom}}_{\beta\varepsilon,\alpha E} =A_{\alpha\beta}\delta(E-\varepsilon)$,
\begin{equation}
c_{\beta\varepsilon,\alpha E} =c'_{\beta\varepsilon,\alpha E} + c^{\mathrm{Hom}}_{\beta\varepsilon,\alpha E}.
\end{equation}
The choice of matrix $\mathbf{A}$ dictates the boundary conditions and normalization of the solution. Here, we choose $A_{\alpha\beta}=\delta_{\alpha\beta}-i\pi \langle \beta \varepsilon | \hat{V}' |\mathbf{\Phi}\rangle \mathbf{f}_{\alpha E}$, resulting in:
\begin{equation}\label{eq:c-}
c_{\beta\varepsilon,\alpha E} =\delta_{\alpha\beta}\,\delta(\varepsilon-E)\,+\,
\frac{\langle \beta \varepsilon | \hat{V}' |\mathbf{\Phi}\rangle \mathbf{f}_{\alpha E}}{E-\varepsilon+i0^+}.
\end{equation}
This formulation corresponds to a state normalized as $\langle\Psi_{\alpha E}|\Psi_{\beta E'}\rangle = \delta_{\alpha \beta}\delta(E-E')$ and fulfilling outgoing boundary conditions~\cite{Newton66,Taylor72}. Replacing this expression in~\eqref{Orig1}, we find an equation for $\mathbf{f}_{\alpha E}$,
\begin{equation}\label{eq:bcoef}
\mathbf{f}_{\alpha E} = \frac{1}{E-\tilde{\mathbf{H}}(E)}\langle \mathbf{\Phi}|\hat{V}'|\alpha E\rangle.
\end{equation}
where we defined the complex effective Hamiltonian $\tilde{\mathbf{H}}(E)$ in the space of localized states as:
\begin{equation}\label{eq:Htilde}
\tilde{\mathbf{H}}(E)=\mathbf{E}+\sum_\beta\int d\varepsilon 
\frac{\langle \mathbf{\Phi}|\hat{V}'|\beta\varepsilon\rangle\langle \beta\varepsilon | \hat{V}' |\mathbf{\Phi}\rangle}{E-\varepsilon+i0^+}.
\end{equation}

The term added to the diagonal matrix $\mathbf{E}$ is the sum of the principal part of the integral, and of the residual,
\begin{equation}\label{eq:Etilde}
\tilde{\mathbf{H}}(E)=\mathbf{E}+\boldsymbol{\Delta}(E)-\frac{i}{2}\boldsymbol{\Gamma}(E)
\end{equation}
where $\boldsymbol{\Delta}(E)$ and  $\boldsymbol{\Gamma}(E)$ are defined as:
\begin{eqnarray}
\boldsymbol{\Delta}(E) &=& \sum_\beta\dashint d\varepsilon 
\frac{\langle \mathbf{\Phi}|\hat{V}'|\beta\varepsilon\rangle\langle \beta\varepsilon | \hat{V}' |\mathbf{\Phi}\rangle}{E-\varepsilon},\\
\boldsymbol{\Gamma}(E) &=& 2\pi\sum_\beta 
\langle \mathbf{\Phi}|\hat{V}'|\beta E\rangle\langle \beta E| \hat{V}' |\mathbf{\Phi}\rangle.\label{GammaMatrix}
\end{eqnarray}
To save space, we introduce the vectors $\left[\mathbf{R}_{\alpha E}\right]_i=\langle \Phi_i | \hat{V}' | \alpha E\rangle$. The expansion coefficients on continuum states are then given by:
\begin{equation}\label{eq:ccoef}
c_{\beta\varepsilon,\alpha E} =\delta_{\alpha\beta}\delta(\varepsilon-E)\,+\,
\frac{\mathbf{R}_{\beta\varepsilon}^\dagger[E-\tilde{\mathbf{H}}(E)]^{-1}\, \mathbf{R}_{\alpha E} }{E-\varepsilon+i0^+}.
\end{equation}

The non-Hermitian effective Hamiltonian $\tilde{\mathbf{H}}(E)$ can be diagonalized, finding a set of left and right eigenvectors,
\begin{equation}
\begin{split}
     \tilde{\mathbf{H}}(\tilde{E}_i)\, \mathbf{u}_{R,i} &= \mathbf{u}_{R,i}\,\tilde{E}_i\\
     \mathbf{u}_{L,i}^\dagger\,\tilde{\mathbf{H}}(\tilde{E}_i)  &= \tilde{E}_i\,\mathbf{u}_{L,i}^\dagger.
\end{split}
\end{equation}
Assuming that both $\boldsymbol{\Delta}(E)$ and $\boldsymbol{\Gamma}(E)$ depend only weakly on the parameter $E$, the effective Hamiltonian is essentially the same when evaluated at any of its eigenvalues, $\tilde{\mathbf{H}}(\tilde{E}_i)\simeq\tilde{\mathbf{H}}(\tilde{E}_j)$ for any $i$ and $j$. This allows us to drop the energy dependence of the effective Hamiltonian, denoting it simply as $\tilde{\mathbf{H}}$, and solve for all the eigenvectors simultaneously,
\begin{equation}
    \tilde{\mathbf{H}}\,\tilde{\mathbf{U}}_{R}=\tilde{\mathbf{U}}_{R}\,\tilde{\mathbf{E}},\qquad \tilde{\mathbf{U}}_{L}^\dagger\,\tilde{\mathbf{H}}=\tilde{\mathbf{E}}\,\tilde{\mathbf{U}}_{L}^\dagger,
\end{equation}
where $\tilde{\mathbf{E}}_{ij}=\delta_{ij}\tilde{E}_i$ is the diagonal matrix of the complex resonance energies, $\tilde{E}_i=\bar{E}_i-i\Gamma_{i}/2$, and $\tilde{\mathbf{U}}_{R/L}$ are the matrices of the right and left eigenvectors, respectively, which, as usual, can now be chosen to be normalized as $\tilde{\mathbf{U}}_{L}^\dagger\tilde{\mathbf{U}}_{R}=\mathbf{1}$.
The expression for $\mathbf{f}_E$ then becomes:
\begin{equation}\label{eq:b2}
\mathbf{f}_{\alpha E} = \tilde{\mathbf{U}}_{R} (E-\tilde{\mathbf{E}})^{-1}\tilde{\mathbf{U}}_{L}^\dagger\mathbf{R}_{\alpha E}. 
\end{equation}
Finally, the expansion coefficient on continuum states are
\begin{equation}\label{eq:c-2}
\begin{split}
c_{\beta\varepsilon,\alpha E} &=\delta_{\alpha,\beta}\,\delta(\varepsilon-E)\,+\\
&+
\frac{\mathbf{R}_{\beta\varepsilon}^\dagger\tilde{\mathbf{U}}_R\,(E-\tilde{\mathbf{E}})^{-1}\,\tilde{\mathbf{U}}_L^\dagger \mathbf{R}_{\alpha E} }{E-\varepsilon+i0^+}.
\end{split}
\end{equation}
To summarize, we have derived expressions for the expansion coefficients of states in a multi-channel continuum, treating all radiative and non-radiative interactions non-perturbatively, while neglecting continuum-continuum radiative coupling.

In the following, we assume that the energy shift of the resonances due to their interaction with the continuum is negligible compared with the energy gaps between the localized states, $\boldsymbol{\Delta}\simeq 0$.
The XUV absorption spectrum $\sigma(\omega_{\textsc{xuv}})$ from the laser-dressed system, initially in the ground state with a large number of photons $n\gg 1$, denoted as $|g,n\rangle$, is proportional to the square module of the total dipole transition amplitude, which is given by the incoherent sum of the square module to all the available continuum channels,
\begin{equation}\label{eq:TASeqn}
    \sigma(\omega_{\textsc{xuv}})\propto \sum_\alpha |\langle g, n|\hat{p}|\psi_{\alpha E}\rangle|^2,
\end{equation}
where $E=E_g+n\omega+\omega_{\textsc{xuv}}$ and $\hat{p}=\hat{\epsilon}\cdot\hat{\vec{p}}$.
Using expression~\eqref{eq:PsiGeneral}, the partial photoionization amplitude $\langle g, n|\hat{p}|\psi_{\alpha E}\rangle$  can be written as
\begin{equation}
\begin{split}
\langle g, n|\hat{p}|\psi_{\alpha E}\rangle &= \langle g, n|\hat{p}|\mathbf{\Phi}\rangle\mathbf{f}_{\alpha E}  +\\
&+\sum_\beta\int d\varepsilon \langle g, n|\hat{p}| \beta\varepsilon \rangle c_{\beta\varepsilon,\alpha E}.
\end{split}
\end{equation}
Assuming that the radiative coupling with the unperturbed non-resonant continuum is constant in the energy region of interest, $\langle g, n | \hat{p} | \alpha E\rangle = \delta_{n,n_\alpha}p_{g\alpha}$, and using the expressions for the bound and continuum coefficients in~\eqref{eq:b2} and~\eqref{eq:c-2}, the matrix element between the ground state and the resonant continuum can be expressed as:
\begin{equation}\label{eq:GeneralizedAmplitude}
\langle g, n | \hat{p} | \psi_{\alpha E}\rangle = p_{g\alpha}+p_{g\alpha}\sum_j\frac{\tilde{q}_j-i
}{2(E-\bar{E}_j)/\Gamma_j+i},
\end{equation}
where 
the complex asymmetry parameter $\tilde{q}_j$ is defined as:
\begin{widetext}
\begin{equation}
    \tilde{q}_j = \frac{2}{p_{g\alpha}\Gamma_j}\left[\left(
p_{gn,\boldsymbol{\Phi}}+\sum_\beta p_{g\beta}\dashint  \frac{\mathbf{R}_{\beta\varepsilon}^\dagger\,d\varepsilon}{E_j-\varepsilon}\right)\tilde{\mathbf{P}}_{j}\mathbf{R}_{\alpha E}-i\pi\sum_\beta \mathbf{R}_{\beta E}^\dagger\tilde{\mathbf{P}}_{j}(p_{g\beta}\mathbf{R}_{\alpha E}-p_{g\alpha}\mathbf{R}_{\beta E})
\right],
\end{equation}
\end{widetext}
with $\tilde{\mathbf{P}}_j=\tilde{\mathbf{U}}_{Rj}\tilde{\mathbf{U}}_{Lj}^\dagger$ being the projector on resonance $j$.
Notice that the transition amplitude~\eqref{eq:GeneralizedAmplitude} is a coherent sum of Fano profiles, one for each autoionizing polariton~\cite{Argenti2013}, generalized to dissipative systems (the $q$ parameter is complex rather than real). The interference between the different terms cannot in general be neglected. This expression, therefore, is essential to correctly fit the complex polaritonic multiplets observed in the experiment, as it was done in~\cite{Harkema2022}, and is the main result of this section.

In the specific case of two AIPs, such as the one examined in the previous section, the system comprises two bound and four continuum components. The bound components, $|\mathbf{\Phi}\rangle = ( |\psi_1\rangle, |\psi_2\rangle )$ (where we use indices $1$ and $2$ for convenience instead of $+$ and $-$), are expressed in terms of $|a\rangle$ and $|b\rangle$ by Eq.~\eqref{eq:Psipm}. The energies of these components depend parametrically on the laser intensity $I$ and angular frequency $\omega$, as detailed in~\eqref{eq:Epm}.
The four continua are $|\alpha(E+2\omega),n-2\rangle$, $|\beta(E+\omega),n-1\rangle$, $|\alpha E, n\rangle$, $|\beta(E-2\omega),n+1\rangle )$.
Using Eq.~\eqref{GammaMatrix}, we can calculate the elements of the matrix $\mathbf{\Gamma}$, which can be written in general as
\begin{equation}
\mathbf{\Gamma}_{ij} = 2\pi\left[A_{ij}  + B_{ij}\sin(2\theta+\phi_{ij})\right].
\end{equation}
The symbols $A_{ij}$, $B_{ij}$, and $\phi_{ij}$ are real parameters, which can be determined using appropriate trigonometric relations.
Using the relationship \eqref{eq:Etilde} we can build and diagonalize the matrix $\mathbf{\tilde{H}} = \mathbf{E} - i \mathbf{\Gamma}/2$, where $\mathbf{E}_{ij}=\delta_{ij}E_i$,
\begin{equation}
 \mathbf{\tilde{E}} =\begin{bmatrix}
 E_+ -i\Gamma_{11}/2& -i\Gamma_{12}/2 \\\\
 -i\Gamma_{21}/2 &  E_- -i\Gamma_{22}/2
 \end{bmatrix},
\end{equation}
which has eigenvalues
\begin{eqnarray}
    E_{\Omega,1/2} &=& \frac{E_1+E_2  -i\left(\Gamma_{11}+\Gamma_{22}\right)/2}{2} \pm\\
    &\pm& \frac{\sqrt{(E_1-E_2-i\left(\Gamma_{11}-\Gamma_{22}\right)/2)^2-\Gamma_{1,2}\Gamma_{2,1}}}{2}.\nonumber
\end{eqnarray}
From these results, we can evaluate the XUV photoabsorption amplitude from~\eqref{eq:GeneralizedAmplitude}, as a function of laser intensity and frequency.
\begin{figure}
    \centering
    \includegraphics[width=\columnwidth]{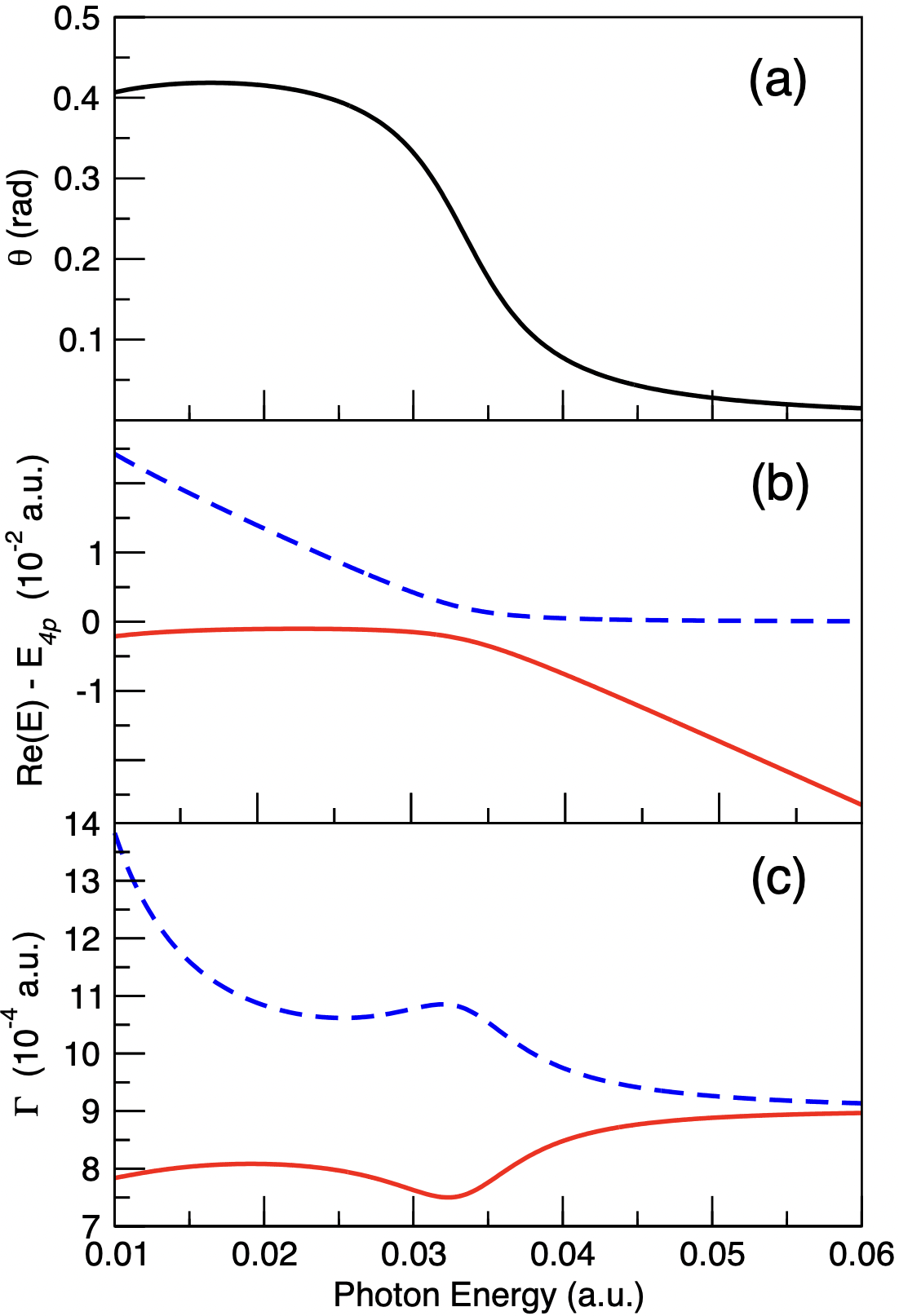}
    \caption{\label{fig:PoleParameters}
    Relevant properties of the autoionizing polaritons, as a function of the dressing-field energy: (a) Mixing angle (see \eqref{eq:Epm}); (b) Real part of the polariton energy with respect to that of the $4p$ bright state; (c) Imaginary part of the polariton energy (i.e., $-\Gamma/2$). See text for details.
    }
\end{figure}
Figure~\ref{fig:PoleParameters} shows the mixing angle and the real and imaginary part of the two polariton complex energy, as a function of the laser frequency. The Light-Induced State (LIS) and bright state come in resonance ($\delta=0$) when $\omega=0.033345$~a.u. The polariton energies (Fig.\,\ref{fig:PoleParameters}\,b) exhibit a clear avoided crossing associated with a change in character of the polaritonic wavefunction, as shown in the rapid transition of the mixing angle from $~90^\circ$ to $~0^\circ$ (Fig.\,\ref{fig:PoleParameters}\,a). 

Observing the imaginary part of the polaritonic energy in Figure\,\ref{fig:PoleParameters}\,c, there is a noticeable dip in the width $\Gamma$ of the lower polariton (red continuous line) at the resonant condition, indicative of its stabilization. Conversely, the width of the upper polariton exhibits a peak at the same frequency, indicative of destabilization. At low frequencies, the width of the upper polariton sharply increases. Although not shown in the figure, the width of the lower polariton also diverges at low frequencies. This behavior is due to the increase in the radiative ionization rate, which is inversely proportional to the laser frequency at constant intensity [see Eq.~\eqref{eq:RadiativeCoupling}]. 

For higher values of the dressing-laser frequency, the LIS and the bright state appear as isolated resonances, with their widths much smaller than their energy separation. Since their lifetime is dominated by the Auger decay rate, which we chose to be identical, their widths converge to the same value.

\begin{figure}
    \centering
    \includegraphics[width=\columnwidth]{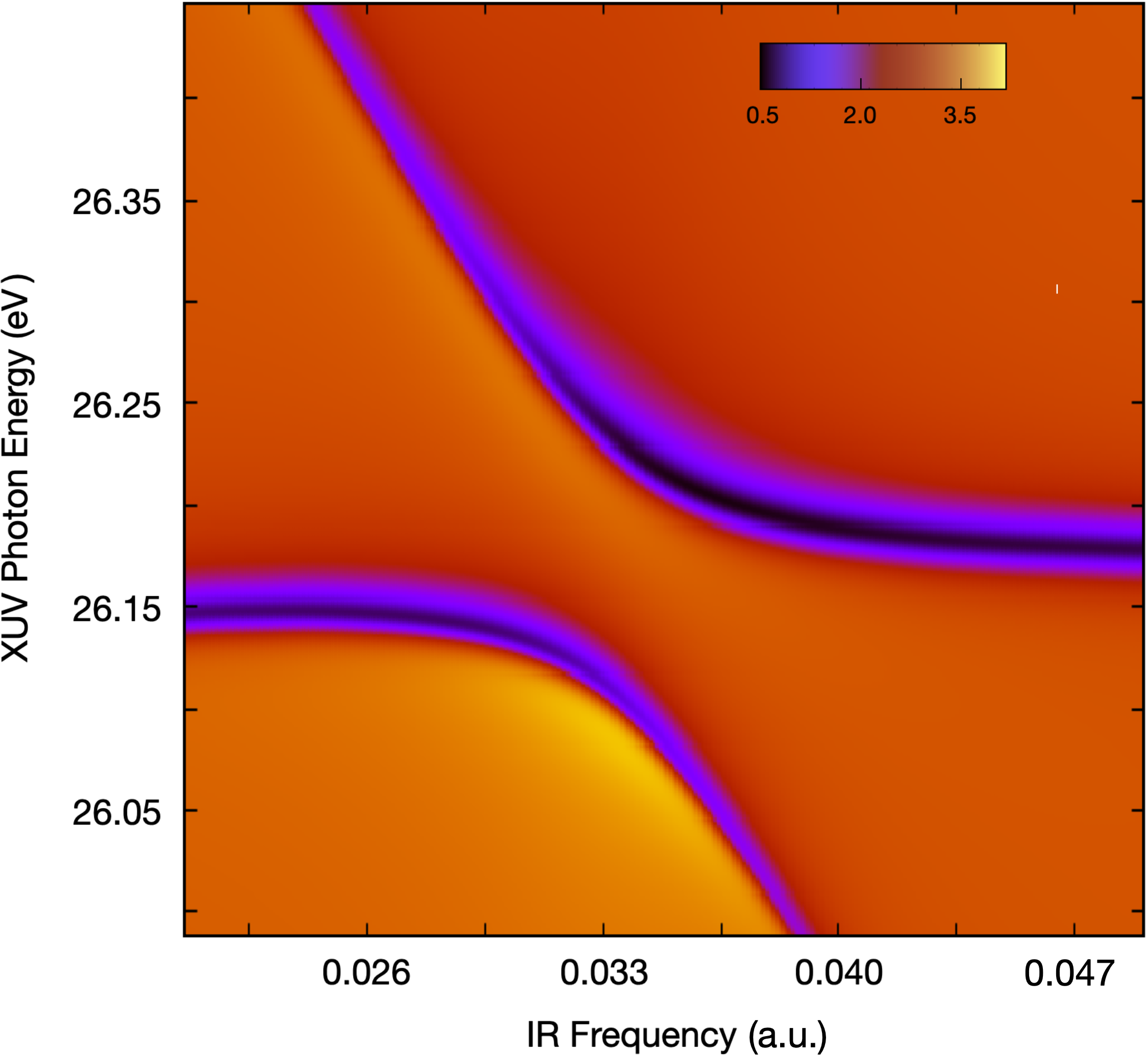}
    \caption{Model-predicted absorption spectrum [in arbitrary units] as a function of IR frequency, based on the formula given in Eq.~\ref{eq:TASeqn}, which accounts for the finite coupling to the non-resonant continuum and the interference between closely-spaced branches.
    \label{fig:fdep}}
\end{figure}

Figure~\ref{fig:fdep} presents the transient absorption spectrum predicted by the model, as a function of the dressing-laser frequency. This spectrum is calculated assuming a field-free $q$ parameter for the two autoionizing states, which reproduces the window character of the $|3s^{-1}4p,n\rangle$ bright state and of the $|3s^{-1}3d,(n-1)\omega\rangle$ LIS, as observed in the experiment~\cite{Harkema2022,Yanez-Pagans2022}. From this spectrum, we can extract both the position and width of the two polaritons, which reproduce those shown in Fig.~\ref{fig:PoleParameters}. 

When these features are well separated, each exhibits a Fano profile. However, when they overlap, the profile is given by the square of the sum of two Fano-like amplitudes, as detaied in~\eqref{eq:GeneralizedAmplitude}, which interfere giving rise to a line shape that cannot be approximated with the sum of two Fano profiles. The absorption spectrum, shown in Figure~\ref{fig:fdep}, takes into account both the finite coupling to the non-resonant continuum and the interference effects between closely-spaced branches.
As anticipated, the avoided crossing of this simplified case closely resembles those reproduced by the \emph{ab initio} calculations, as shown for example in Fig.~\ref{ESBasis}

\section{Conclusions}\label{sec:conclusion}
In this work, we have used~\emph{ab initio} close-coupling simulations to reproduce the resonant attosecond transient absorption spectrum in argon, in the presence of a strong radiative coupling between bright and dark autoionizing states. On the methodological side, we show how the configuration space can be reduced to a small set of essential states, greatly accelerating the numerical solution of the time-dependent Schrödinger equation. The essential-state approach also facilitates the identification of the crucial states that should be included in a realistic modeling of the system. The results of our simulations highlight several phenomena unique to the interplay between Auger decay, Autler-Townes splitting, and radiative ionization of autoionizing states, namely: the stabilization and destabilization of autoionizing states, due to the destructive or constructive interference between radiative and non-radiative decay paths to the ionization continuum, the avoided crossing between window resonances, and the complex profile of overlapping bright and light-induced states. To interpret this phenomenology, we introduced an analytical model that integrates the Jaynes-Cummings formalism with Fano's formalism to describe the optical response of laser-dressed autoionizing states. In this model, the autoionizing states of Fano's formalism are replaced by autoionizing polaritons, entangled states of matter and radiation. This model successfully quantifies the splitting of strongly coupled autoionizing resonances as a function of the driving laser parameters, and explains how competing radiative and Auger ionization pathways can result in the stabilization of an autoionizing polariton. Finally, our model provides analytical expressions for the line shapes of overlapping laser-dressed autoionizing states in transient absorption spectra, which are essential to extract the parameters of the laser-dressed resonances from the experimental spectrum.

\section{Acknowledgements}
We gratefully acknowledge the support received for this work. LA and SM thank the National Science Foundation's Theoretical AMO Physics program for their support through grants No.~1607588 and No.~1912507. EL acknowledges support from the Swedish Research Council under Grant No.~2020-03315. The authors are grateful for the computational resources provided by the UCF ARCC. Additionally, AS, NH, and SP acknowledge the support from the U.S. Department of Energy, Office of Science, Basic Energy Sciences, under Award No.~DE-SC0018251.


%

\end{document}